\documentclass[timestamp, acmtog]{acmart}

\usepackage{booktabs} % For formal tables
\usepackage[caption=false]{subfig}
\usepackage{enumitem}
\usepackage{cleveref}

\def\G{\mathcal{G}}
\def\E{\mathcal{E}}
\def\V{\mathcal{V}}
\def\tP{\tilde{\Phi}_0}

\newcommand{\myparagraph}[1]{{\vspace{2mm} \noindent \textbf{\textit{#1}}}}

\setcitestyle{square}
\hypersetup{draft}

\usepackage[ruled]{algorithm2e} % For algorithms

\SetAlFnt{\small}
\SetAlCapFnt{\small}
\SetAlCapNameFnt{\small}
\SetAlCapHSkip{0pt}
\IncMargin{-\parindent}

\setcopyright{usgovmixed}
\acmDOI{000001.000001_2}

\begin{document}

%\includepdf[pages=1-last]{real.pdf}

% Title portion
\title{Latent Space Representation for Shape Analysis and Learning}

\author{Ruqi Huang}
\affiliation{\institution{Ecole Polytechnique}}
\email{rqhuang88@gmail.com}

\author{Panos Achlioptas}
\affiliation{\institution{Stanford University}}
\email{optas@stanford.edu}

\author{Leonidas Guibas}
\affiliation{\institution{Stanford University}}
\email{guibas@cs.stanford.edu}

\author{Maks Ovsjanikov}
\affiliation{\institution{Ecole Polytechnique}}
\email{movsjanka@gmail.com}
\renewcommand\shortauthors{}

\begin{abstract}
  We propose a novel shape representation useful for analyzing and processing shape collections, as well for a variety
  of learning and inference tasks. Unlike most approaches that capture variability in a collection by using a template
  model or a base shape, we show that it is possible to construct a full shape representation by using the \emph{latent
    space} induced by a functional map network, allowing us to represent shapes in the context of a collection without
  the bias induced by selecting a template shape.  Key to our construction is a novel analysis of latent
    functional spaces, which shows that after proper regularization they can be endowed with a natural geometric structure,
    giving rise to a well-defined, stable and fully informative shape representation. We demonstrate the utility of
  our representation in shape analysis tasks, such as highlighting the most distorted shape parts in a collection or
  separating variability modes between shape classes. We further exploit our representation in learning applications by
  showing how it can naturally be used within deep learning and convolutional neural networks for shape classification
  or reconstruction, significantly outperforming existing point-based techniques.
\end{abstract}

%
% The code below should be generated by the tool at
% http://dl.acm.org/ccs.cfm
% Please copy and paste the code instead of the example below.
%

%
% End generated code
%


\maketitle

\section{Introduction}\label{sec:intro}

Detecting, quantifying and analyzing variability in shape collections is a fundamental task in computer graphics and geometry processing, with applications across multiple domains, including in statistical shape analysis \cite{scape,faust,hasler09}, shape exploration \cite{kim2012exploring,Rustamov2013,kleiman2015shed}, shape correspondence \cite{huang2014functional} and co-segmentation \cite{wang2012active}. A key question that arises in all techniques for extracting variability is the choice of the right \emph{shape representation}, which can reveal the structure of each shape in the context of the collection while also being compact and easy to manipulate, enabling efficient shape analysis and processing.

The majority of existing techniques dedicated to extracting variability in a collection are based on first selecting a template (or base) shape and considering the changes on all other shapes with respect to this template --- this is the standard practice in medical domains where the reference shape is often referred to as an ``atlas'' (e.g., in brain anatomy) \cite{grenander1998computational}. In computer graphics this approach is common both in shape reconstruction and in statistical shape analysis \cite{scape,faust,hasler09}, but also in shape exploration (e.g., \cite{kim2012exploring,ovsjanikov2011exploration,kim2013learning,Rustamov2013} among many others) where the template is often constructed by either simplifying some fixed base shape or by using shape abstractions derived from collections of parts and their relations.

Although easy and intuitive, template-based shape exploration and analysis has obvious limitations when shape variability is large and no single prototype adequately models all given shapes. But even in settings of more modest variation, there are significant limitations: first, the choice of the template can significantly affect the results in terms of the types of variability that is detected and highlighted. Second, considering the variability with respect to a fixed base shape can make it difficult to reveal cross-class variability that becomes apparent only when comparing all pairs of shapes in the collection. Third, even when the base shape is given, the exact choice of encoding for the variability remains crucial. For example, while simple techniques based on the displacement of each template vertex might be relevant for reconstruction and statistical shape analysis, their use is very limited in the context of learning since, as they are not invariant to even the basic rigid motions.

This question of shape representation has also become particularly important with the advent of powerful techniques based on deep learning and convolutional neural networks. Although very successful in image analysis, their adoption for shape processing has so far been relatively limited due to representational differences. Common 3D representations such as meshes or point clouds are irregular, unlike the regular grids defining 2D images, making it challenging to define notions such as convolution or to encode basic 3D invariances. While some significant progress has been made in this direction in the past few years (see e.g., \cite{,maron2017convolutional} and \cite{bronstein2017geometric} for an overview), the question of defining a representation that is at once invariant, compact and well-suited for learning remains open.

In this paper we present a novel approach to encoding shapes in the context of a collection that helps overcome many of the above limitations. Specifically, starting from a collection of shapes with some soft (functional) maps between them, we show how consistent latent spaces that have previously been used for improving map quality can also be exploited to reveal the geometric variability in the collection, without relying on a base or template shape (e.g., in Figure~\ref{fig:two_layer}, our approach highlights the regions that are distinctive between the cats and lions), or assuming a particular (e.g., star-shaped) topology of the functional map network. Our approach is based on a novel analysis of latent spaces, which demonstrates that after proper regularization they can be endowed with natural, unbiased geometric structure. We then show that, although our latent shape is a dual object that need not correspond to a real shape in 3D, it can be used together with the notion of shape differences introduced in \cite{Rustamov2013} to construct a representation for each shape in the collection, but without relying on a fixed base shape as done in that work. Moreover, we show how the \emph{algebraic nature} of this representation can be exploited to detect detailed information about differences between shape classes, perform (possibly partial) shape analogies, and analyze shapes across different modalities.

\vspace{4mm}\noindent\textbf{Contributions.}
To summarize, our main contributions are:
\begin{itemize}
\item We describe how latent functional spaces can be endowed with natural geometric (metric and measure) structure, giving rise, for the first time, to a well-defined notion of a ``latent shape'' that characterizes a shape collection.
    \item We define shape differences between real and latent shapes and show how such differences lead to a shape representation, that can be used for detailed shape analysis without assuming a particular topology of the map network.
    \item We provide tools for a nuanced understanding of shape viability, including the separation of the different types of variability present within and across shape sub-collections.
    \item We demonstrate that our new representation supports deep learning techniques, including CNNs, for both analysis and synthesis, leading to improved results over baseline methods.
\end{itemize}

\begin{figure*}[h]
  \centering
  \includegraphics[width=0.9\linewidth]{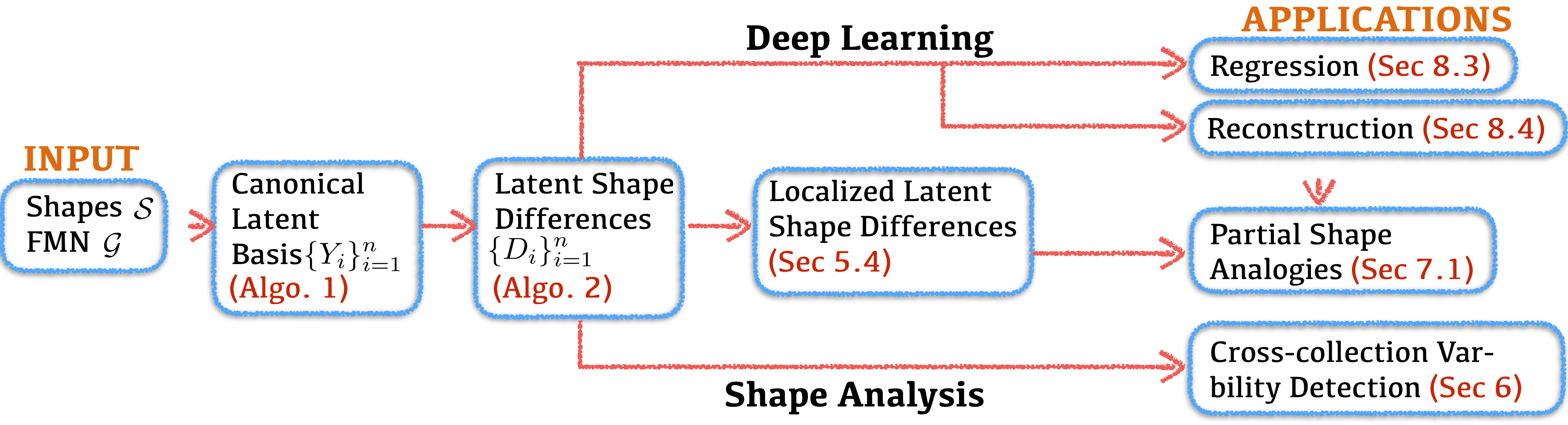}
        \caption{\label{fig:flow} We start with a collection of shapes and a set of functional maps among them, from which we extract a latent shape $S_0$. And then we construct for each shape a canonical latent basis, and represent each shape with a pair of operators (matrices), which are based on the latent basis. Finally, we demonstrate various applications of the \emph{latent representations} in shape analysis and deep learning on 3D shapes.   }
\end{figure*}

%!TEX root = main.tex
\section{Related Work}\label{sec:related}

\noindent\textbf{Template-based shape analysis and exploration.}
Analyzing shape collections by variability around a template shape has a rich and vast history going
back to D'Arcy Thompson's classic ``On growth and form'' \cite{thompson1942growth}, which has
inspired Kendall's shape space theory \cite{kendall1989survey} and pattern theory
formalized by Grenader and commonly used in computational anatomy \cite{grenander1998computational},
where templates are often referred to as \emph{atlases}.

In Computer Graphics, shape spaces based on template variation are ubiquitous in statistical shape
analysis, e.g. for defining 3D morphable models \cite{blanz1999morphable,allen2003space}, especially
for capturing variability in human body and pose, e.g. \cite{scape,hasler09,faust} among many others.

Shape templates are also commonly used for exploring shape collections
\cite{ovsjanikov2011exploration,kim2012exploring}. Although in most cases the presence of a shape
template is assumed to be given \emph{a priori}, simultaneous template construction and fitting
techniques have been used for both reconstruction \cite{wand2007,wand2009efficient,tong2012scanning}
and exploration \cite{kim2013learning}, among many others.

While pervasive, template-based methods also have a well-known limitation in that the choice of the
template model can introduce bias in the kinds of variability that are revealed. Common
selection techniques include using a particular (median) shape in a collection that is as close as
possible to a centroid, or constructing a new template shape by pointwise averaging (e.g.,
\cite{joshi2004unbiased}).

Our approach avoids the construction of a explicit template shape, and replaces it with an implicit
template obtained via the analysis of latent functional space, which both removes the bias in the
template shape selection and also avoids the expensive geometric (embedded 3D shape) template construction.

\vspace{1mm}

\noindent\textbf{Shape Analysis with functional maps.}
Our approach takes as input a collection of shapes with soft (functional) maps between them. In
this, we follow the recent line of work on shape analysis with soft maps, similar to
\cite{solomon2012soft,kim2012exploring,Rustamov2013}. Namely, we use the formalism of
\emph{functional maps} introduced originally in \cite{Functional} and extended significantly in
follow-up works, including \cite{kovnatsky2013coupled,huang2014functional} among others (see
\cite{ovsjanikov2017computing} for a recent overview).

Although originally proposed as a computational tool for shape matching, follow-up works have also
shown its utility in shape analysis and exploration, starting with map visualization
\cite{Ovsjanikov2013}, detection and encoding of \emph{shape differences} \cite{Rustamov2013}, and
co-segmentation and co-analysis \cite{huang2014functional} among others.  The advantage of these
techniques is that they only require approximate functional maps, which are much easier to compute
than precise (point-to-point) correspondences. Nevertheless, existing methods such as
\cite{Ovsjanikov2013,Rustamov2013} also follow the spirit of template-based techniques and assume
the presence of a single \emph{base} shape with respect to which variability is captured. A recent
method introduced in \cite{Huang2017} has tried to lift this assumption but is still restricted to
revealing global variability within a single collection. We extend these techniques first by
proposing a template-free analysis and exploration framework using functional maps and second by
proposing techniques for detecting and highlighting \emph{cross}-collection variability, and finally
by defining a compact shape representation that is suitable for learning.

\vspace{1mm}

\noindent\textbf{Latent functional spaces.}
A key building block in our approach is the use of so-called latent functional spaces, which are
closely related to map synchronization \cite{wang2013exact} and which have been used for computing
\emph{consistent functional maps} in shape and image collections
\cite{huang2014functional,Wang2013,wang2014unsupervised}. One of our key contributions is to show
that in addition to providing a powerful computational method for map inference,
latent functional spaces also allow to reveal variability in shape collections and also to define a
compact and informative shape representation.

\vspace{1mm}

\noindent\textbf{Shape representations for learning.}
One of our key applications is to show how the shape representation obtained via the latent
functional spaces can be naturally used in the context of \emph{supervised learning} applications, and
especially enable the use of convolutional neural networks for shape regression and classification.

In this, our work is related to the recent techniques aimed at applying deep learning methods to
shape analysis. One of the main challenges is defining a meaningful notion of convolution, while
ensuring invariance to basic transformations, such as rigid motions. Several techniques have recently
been proposed based on e.g., Geometry Images \cite{sinha2016deep}, Volumetric
\cite{maturana2015voxnet,wang2017cnn}, point-based \cite{qi2016_pointnet} and multi-view approaches
\cite{Su_volumetric}, as well as, more recently intrinsic techniques that adapt convolution to
curved surfaces \cite{masci2015geodesic,boscaini2016learning} (see also
\cite{bronstein2017geometric} for an  overview), and even via toric covers
\cite{maron2017convolutional} among many others.

Despite this tremendous progress in the last few years, defining a shape representation that can
naturally support convolution operations, is compact, invariant to the desired class of
transformations (e.g., rigid motions) and not limited to a particular topology, remains a
challenge. As we show below, our representation is well-suited for learning applications, and
especially for revealing subtle geometric information regarding the shape structure.

\vspace{1mm}

\noindent\textbf{Shape processing in latent representations.}
Finally, our work is also related to recent techniques that construct latent spaces for 
representing 3D shapes, especially those based on learning.
For instance, \cite{3dgan} combine a 3D-CNN with a Generative Adversarial Network (GAN) to first learn the latent
space of 3D shapes. Given the latent space, they regress an image feature learned via a 2D-CNN to
the latent space to recover the underlying geometry. \cite{girdhar1eccv} follow a similar strategy
but use a voxel-based AutoEncoder (AE) instead of a GAN for learning the latent representation.
\cite{achlioptas2017latent_pc} introduced an AE operating on 3D point-clouds to produce a latent
space which is further exploited by a GAN for point-cloud synthesis. In a similar manner,
\cite{LiXCYZG17} developed a recursive neural net to map 3D part-layouts to a latent space at which
a GAN operates to create novel shapes with various part-hierarchies.

Differently from our representation via latent space analysis, these learned embeddings
represent shapes {\em as points} in some high-dimensional space and rarely give access to regions or parts
of the 3D shapes associated with or responsible for the shape variability. On the other hand, we
represent shapes in a collection as {\em linear operators}, stored as matrices, which not only enables a
meaningful notion of convolution but also allows us to recover \emph{explanations} for differences
and variability in terms of highlighted shape pats.

%!TEX root = main.tex
\section{Overview}
\label{sec:overview}

The rest of the paper is organized as follows: in Section \ref{sec:prelims} we describe the problem
setting, the main goals and notations used below. Section \ref{sec:geometry} provides the theoretical foundation for our method. 

In particular, we characterize the geometric structure of latent shapes in Section \ref{sec:latentShape} and define our shape representation based on shape differences with respect to latent shapes in Section \ref{sec:lsd}. We then describe the two key applications: extracting variability in shape collections (Section \ref{sec:comparison}) and using our representation for 3D deep learning (Section \ref{sec:learning}). Finally, we 
show qualitative and quantitative results obtained using our methods in Section \ref{sec:experimental_result}.

\section{Preliminaries, Notation and Problem Setup}
\label{sec:prelims}

Throughout our work, we assume that we are given a collection of related 3D shapes and a set of functional maps~\cite{Functional} among some shape pairs.
Our main goal is to develop a theoretical foundation for a novel representation for the shapes in the collection, and to show how this representation can be effectively used in practical applications.

Specifically, we assume as input a set of shapes $\mathcal{S} = \{S_k\}_{k = 1}^n$ and functional maps $C_{ij}$, which map real-valued functions between some pairs of shapes $S_i,S_j$. The functional maps can either be induced by point-wise correspondences, or, can be obtained via an optimization procedure, as described, e.g., in \cite{ovsjanikov2017computing}. Let $L_i, M_i$ be the stiffness matrix and the area matrix of these shapes, which encode respectively the metric and the measure information. The Laplace-Beltrami operator (LBO) is classically discretized as $M_i^{-1}L_i$~\cite{meyer2003discrete}.  We let $\Lambda_i$ be the diagonal matrix storing the $k$ smallest eigenvalues of the LBO of shape $i$, and $\Phi_i$ the matrix storing the corresponding eigenvectors. Following previous works, we assume that functional maps are given in the reduced eigenbasis and can be thought of as matrices of size $k \times k$.

The functional map network (FMN) on $\mathcal{S}$ is a graph $\G = (\V, \E)$, where the $i-$th vertex in $\V$ corresponds to the functional space on $S_i$, and the edge $(i, j)\in \E$ if we are given a functional map $C_{ij}$. We assume that this network is symmetric ($(i,j) \in \E$ if and only if $(j,i) \in \E$) and is \emph{connected} so that there exists at least one path consisting of the edges in $\E$ between any pair of vertices in $\V$.

\vspace{-2mm}
\paragraph{Shape Differences} Our shape representation is based on the shape differences introduced in~\cite{Rustamov2013}, which characterize shape deformations by encoding the changes in inner products of functions. Namely, given shapes $S_i, S_j$ and a functional map $C_{ij}$ in the reduced basis, the authors introduce the area-based and the conformal shape differences $D^A, D^C$:
\vspace{-2mm}
\begin{align}
	D_{ij}^A = C_{ij}^T C_{ij}, \\
	D_{ij}^C = \Lambda_i^{+} C_{ij}^T \Lambda_j C_{ij},
\end{align}
where $^{+}$ is the Moore-Penrose pseudo-inverse. Intuitively $D_{ij}$ is a linear operator, which once again, can be represented as a matrix of size $k\times k$, and which encodes the difference or distortion induced by a map $C_{ij}$ (see Figure 2 and Eq.(4) in \cite{Rustamov2013}). 

The key limitation of shape difference operators for shape collection analysis, is that they require a choice of a base shape $S_i$ and consider only \emph{directional} changes, from shape $S_i$ to other shapes, making it impossible to use them given an arbitrary (non star-shaped) FMN. Thus, one of our goals is to extend this construction to the case of shape collections without assuming a fixed base shape. We achieve this by exploiting the formalism of latent functional bases~\cite{Wang2013}, which has been proposed for improving the consistency of functional maps.

\vspace{-2mm}
\paragraph{Latent Spaces} Given a FMN, the authors of ~\cite{Wang2013} propose to extract a set of \emph{consistent latent bases} $Y_i$ on $S_i$ such that $C_{ij}Y_i \approx Y_j, \forall i, j$, and use them to refine the quality (consistency) of functional maps.  The latent bases $Y_i$ can be thought of as functions on $S_i$, or as functional maps from some \emph{latent shape} to each shape $S_i$. Then, a map from $S_i$ to $S_j$ can be factored into a map from $i$ to the latent shape and then to $j$ via: $C_{ij} \approx Y_{j} Y_{i}^{-1}$. While useful as a tool for improving functional maps, the exact structure of latent shapes is still not fully understood, and they have so far not been used for \emph{representing} shapes in a collection.

In our work we first show how latent shapes can be endowed with geometric structure, and be made more stable, through an extra regularization, and then define a latent space shape representation.

%!TEX root = main.tex

\section{Latent Representation}
\label{sec:geometry}

\subsection{Canonical Latent Basis and Latent Shape}\label{sec:latentShape}
Our first key observation is that the latent shape plays the role of an ``average shape'' in
analyzing shape collections -- a shape-like object that represents the entire collection, and which
can be endowed with a natural geometric structure. Crucially, unlike existing approaches, for
example in computational anatomy~\cite{younes2010shapes} that consider building templates or average
shapes, we characterize the latent shape \emph{directly in the functional domain}, without
attempting to embed it in the ambient space.

\begin{algorithm}[t!]
\DontPrintSemicolon
\SetKwData{Left}{left}\SetKwData{This}{this}\SetKwData{Up}{up}
\SetKwFunction{Union}{Union}\SetKwFunction{FindCompress}{FindCompress}
\SetKwInOut{Input}{input}\SetKwInOut{Output}{output}
\Input{A set of consistent latent basis $\{Y_i\}_{i = 1}^n$ learned from a shape collection $S$ and associated FMN $\G$. The eigenbasis $\Phi_i$, eigenvalues $\Lambda_i$ on each $S_i$.  }
\Output{A set of canonical consistent latent basis $\{\tilde{Y}_i\}_{i = 1}^n$, and the eigenbasis $\Phi_0$ and the spectrum $\Lambda_0$, for the latent shape.}
\begin{enumerate}[label={(\arabic{enumi})},ref={Step \arabic{enumi}},leftmargin=*]
	\item Compute the eigen-decomposition of $E = \sum_i Y_i^T \Lambda_i Y_i$ so that $E U = U \Lambda$ and let $\tilde{Y}_i = Y_i U$.\\
	\item Let $\Phi_0 = \Psi_i \tilde{Y}_i$ for an arbitrary $i$, and $\Lambda_0 = \Lambda$ from the previous step.
\end{enumerate}
\caption{Computing a Canonical Consistent Latent Basis}
\label{alg:canonical_CLB}
\end{algorithm}

The following theorem establishes the connection between the consistent latent basis and the
geometry of the latent shape, while at the same time highlighting the limitations of the previously
used approaches for constructing latent bases:

\begin{theorem}\label{thm:latentshape}
  Given a collection of discrete 3D shapes in $1$-$1$ vertex correspondence and sharing the same
  mesh connectivity, and a consistent FMN $\G$, in which the functional maps are represented in the
  eigenbasis $\Phi_i$ on each $S_i$.  Let $Y_i$ be the consistent latent basis satisfying the
  conditions: $C_{ij}Y_i = Y_j, \forall (i, j)\in \G, s.t. \sum_i Y_i^T Y_i = I$, and
  $\sum_i Y_i^T \Lambda_i Y_i = \Lambda$, where $\Lambda$ is a diagonal matrix.  Then, the
  eigenbasis $\Phi_0$ of the latent shape whose metric and measure are given by
  $L = \frac{1}{n}\sum_i L_i, M = \frac{1}{n} \sum_i M_i$, i.e.  $L \Phi_0 = M \Phi_0 \Lambda_0,$ can be recovered as
  $\Phi_0 = \Phi_i Y_i$ for any $i$.
\end{theorem}

This theorem suggests that the consistent latent basis carries information about the ``average''
geometry in the collection, given, in the full basis, by the average metric and measure matrices. 

\begin{figure}[t!]
  \centering
  \includegraphics[width=1\linewidth]{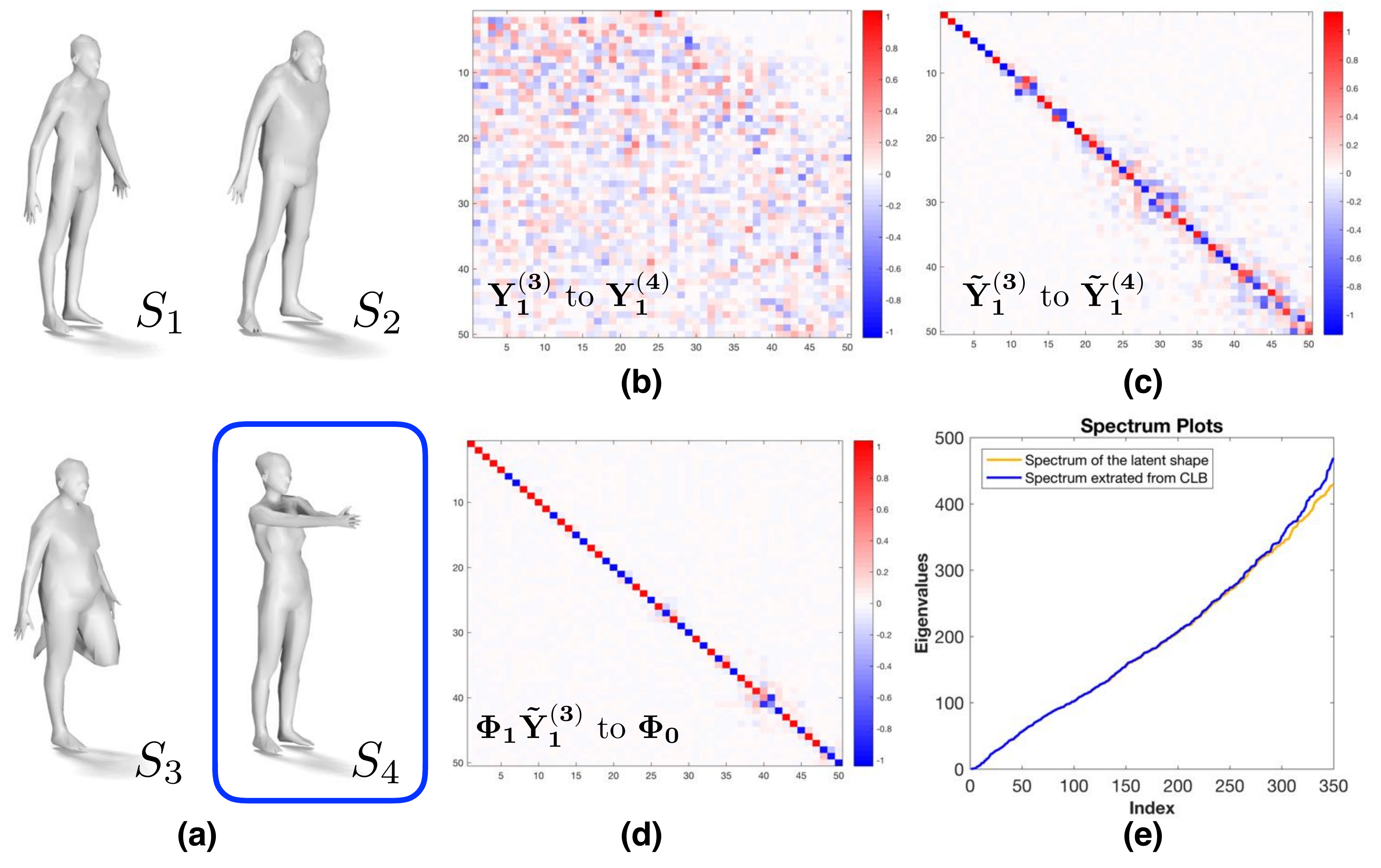}
  \caption{\label{fig:average_shape} (a) input shapes, where $S_4$ is an additional shape; (b) the
    transformation matrix between the standard latent basis $S_1$ computed with and without $S_4$;
    (c) the same transformation matrix but between canonical latent bases. (d) The transformation
    matrix between the computed latent basis and the theoretical ground-truth stated in
    Theorem~\ref{thm:latentshape} when expressing functional maps in a reduced basis (e) the computed
    spectrum and the theoretical ground-truth. (see the text for details)\vspace{-2mm}.}
\end{figure}

\paragraph{\textbf{Role of Proper Regularization}}
Note that previous approaches for constructing the latent basis, such as \cite{Wang2013} proposed to
compute the latent basis by solving the optimization problem
$\min_Y \|C_{ij} Y_i - Y_j\| s.t. \sum_i Y_i^T Y_i = I.$ Geometrically, and in light of Theorem
\ref{thm:latentshape}, this corresponds to only averaging \emph{the measure} of the shapes, which
leads to metric ambiguity. This can result in significant instabilities in the extraction of the
latent basis. We demonstrate this effect in Figure~\ref{fig:average_shape}.  Namely, given a shape
collection $S_1, S_2, S_3$ and an additional shape $S_4$, we compared the CLB $Y_1^{(3)}$ on $S_1$,
with respect to the original shape collection, and $Y_1^{(4)}$ recomputed with all $4$ shapes.
Figure~\ref{fig:average_shape}(b) depicts the change of basis matrix between these two settings,
which has noisy off-diagonal entries, suggesting that the latent shape is significantly perturbed.

To overcome this instability, we propose to construct a \emph{canonical} latent basis by introducing
an extra normalization which forces $\sum_i Y_i^T \Lambda_i Y_i$ to be a diagonal matrix, and which
corresponds in Theorem \ref{thm:latentshape} to averaging the metric on the latent shape. With this
additional normalization, the change of basis matrix between latent bases with and without shape
$S_4$ shown in Figure~\ref{fig:average_shape}(c) is much closer to a diagonal one than in
Figure~\ref{fig:average_shape}(b). The details of this construction are given in
Algorithm~\ref{alg:canonical_CLB}.

Now, the extra normalization incorporates the metric information, therefore the latent shape can be thought of as a well-defined shape. 
In general, a shape with the average metric and measure does not admit an embedding in
$\mathbb{R}^3$, but as we will soon show, this construction carries rich geometric information useful
for shape processing. 

Let us stress that Theorem~\ref{thm:latentshape} is of purely theoretical interest. However, Algorithm~\ref{alg:canonical_CLB} can be implemented in practice, without assuming access to the full basis or exact consistent maps. 
In Figure~\ref{fig:average_shape} (d), we also show the proximity between the eigenbasis/spectrum of
the latent shape recovered from functional maps in the reduced basis and the theoretical ground truth.
Namely, Figure~\ref{fig:average_shape}(d) shows the transformation matrix between the first $50$
computed eigenbasis, when functional maps are represented in a reduced basis of size $400$ and the
theoretical ground-truth, given by the exact averaging of the metric and measure. At the same time,
Figure~\ref{fig:average_shape}(e) shows the eigenvalues in the two cases.
Hereafter, we always use the \emph{canonical latent basis} in all the formulations and applications,
and denote it by $Y_i$ to simplify notation.

\myparagraph{Computing canonical latent basis in practice} Computing the consistent latent basis with the framework of~\cite{Wang2013} involves an eigen-decomposition of a possibly large, block-wise sparse matrix, whose size depends on the number of shapes and the dimensionality of functional maps. In order to gain scalability, in practice we first sample a subset $S_{\text{ini}}$ of shapes with which we compute the canonical latent basis, and then for each shape $S_k$ outside the subset, we search for its nearest neighbor, $S_i$, in $S_{\text{ini}}$ using the Shape-DNA descriptor. Finally, we push the latent basis from $S_i$ to $S_k$ via the functional map $C_{ik}$, namely, $Y_k = C_{ik}Y_i$. In particular, this scheme not only improves the scalability of the computation of our latent shape representation, but also allows to avoid recomputing the latent basis for each new shape. 
\subsection{Shapes as Latent Shape Differences}\label{sec:lsd}
Although the canonical CLB reduces the instability present in the previous basis construction, the
latent bases $Y_i$ unfortunately still cannot be used to represent each shape $S_i$ in the
collection. The main reason is that $Y_i$ is expressed in the eigenbasis $\Phi_i$ of shape $i$, and
therefore, one cannot compare, for example $Y_i$ with $Y_j$, which is fundamental in both shape
analysis and learning applications.

Instead, we build our shape representation by defining the \emph{latent shape differences}, which
are linear operators acting on the function space of the latent shape, and which, as such, are
independent of the basis on each shape.

Namely, we assume the spectrum of the latent shape, arising from step 3. of the procedure described in
Algorithm~\ref{alg:canonical_CLB}, is denoted by $\Lambda_0$.  Then, following the formulation
of~\cite{Rustamov2013}, we define the area-based and conformal latent shape differences as:
\begin{align}
D_i^A &= Y_i^T Y_i\label{equ:area_latent} \\
D_i^C &= \Lambda_0^+ Y_i^T\Lambda_i Y_i,
\end{align} 
The final procedure for extracting these operators from a given collection is summarized in
Algorithm \ref{alg:latentdiffs}.
\begin{algorithm}[t!]
\DontPrintSemicolon
\SetKwData{Left}{left}\SetKwData{This}{this}\SetKwData{Up}{up}
\SetKwFunction{Union}{Union}\SetKwFunction{FindCompress}{FindCompress}
\SetKwInOut{Input}{input}\SetKwInOut{Output}{output}
\Input{Shape collection $S$ and associated FMN $\G$. The eigenbasis basis $\Phi_i$ and spectrum $\Lambda_i$ on each $S_i$. }
\Output{A pair of latent shape differences for each shape $i$: area-based $\{D^A_i\}_{i = 1}^n$ and conformal $\{D^C_i\}_{i = 1}^n$.}
\begin{enumerate}[label={(\arabic{enumi})},ref={Step \arabic{enumi}},leftmargin=*]
 \item Compute the CLB $\{Y_i\}_{i=1}^n$ with respect to $S$ and $\G$ via the framework of~\cite{Wang2013}.\\
 \item Compute the canonical CLB $\{\tilde{Y}_i\}_{i=1}^n$ and diagonal matrix with the spectrum of the latent shape, $\Lambda_0$ using Algorithm~\ref{alg:canonical_CLB}.\\
 \item Construct $D_i^A = \tilde{Y}_i^T \tilde{Y}_i$, and respectively $D_i^C = \Lambda_0^+ \tilde{Y}_i^T \Lambda_i
   \tilde{Y}_i$.
\end{enumerate}
\caption{Construction of Latent Shape Difference Operators}
\label{alg:latentdiffs}
\end{algorithm}

The main insight of our work is that the latent shape differences provide a compact and extremely
versatile representation for each shape in a collection as a pair of small-sized matrices, which
enjoy several nice theoretical properties, and enable a number of novel applications in analysis and learning.

\subsection{Properties of the Latent Shape Differences}
\label{subsec:lssd_properties}
Given a shape collection with the associated functional map network, the latent space shape
differences (LSSDs) provide a representation of each shape as a pair of matrices whose size is
controlled by the size of the latent basis. In this work, we argue that this representantion enables
a number of novel applications and lifts fundamental restrictions of previous approaches. In
particular, LSSDs inherit some of the most attractive properties of shape differences, such as their
compactness and informativeness, while avoiding their shortcomings. Below we summarize the main
properties of this representation.

\myparagraph{Invariance:} LSSDs provide a representation that is invariant to rigid (and more
generally isometric) shape transformations. In the context of learning, this is especially important
as it will allow us to do inference in a pose-invariant way.

\myparagraph{Flexibility:} computing LSSDs only requires the knowledge of functional maps and places
no restriction on the shape discretization. For example, they can accomondate collections of shapes
with different number of vertices, or even with different modalities such as point-clouds and
meshes.

\myparagraph{Informativeness:} LSSDs fully encode the intrinsic geometry of each shape in the
collection in a compact way. Indeed, it follows from Theorem~\ref{thm:latentshape} that in
the presence of full information, given the FMN of a collection of shapes $\mathcal{S}$, the
spectrum $\Lambda_0$ of the latent shape, and $D_i^A, D_i^C$ for each shape in $\mathcal{S}$, one
can recover the intrinsic geometry for each $S_i$, i.e., the area and stiffness matrices $M_i, L_i$,
which, in turn, fully determines the edge lengths \cite{zeng2012discrete}.

\myparagraph{Functoriality:} if we interpret each $Y_i$ as the functional map associating the latent
shape to $S_i$, it follows from the functoriality property in~\cite{Rustamov2013} that
\[D_{ij} = Y_i D_i^{-1} D_j Y_i^{-1}, \]
where $D_{ij}$ is the shape difference between $S_i$ and $S_j$. Thus, LSSDs not only encode the
difference of each shape to the latent shape but also allow to factor the difference between each
pair of shapes, via the canonical latent basis.

\myparagraph{Algebraic nature:} LSSDs are linear functional operators on the latent shape. As such,
they can be represented as small matrices and manipulated using standard numerical
linear algebraic tools, in practice. Moreover, they provide detailed (localized) information about the shape
geometry. As we show below, this allows us to extract \emph{partial} information to compare and
reconstruct shape parts, in contrast to purely global shape descriptors.

\myparagraph{Base-shape independence:} Crucially, unlike the original shape differences, which rely
on the choice of a specific base shape, which can lead to biased results, and requires a star-shaped
map network, LSSDs are extracted from the \emph{entire} input functional map network, regardless of
its topology. This is especially true due to our novel regularization, which leads to a latent shape,
endowed with canonical geometric structure. Let us note that theoretically, in the presence of full
information and an \emph{a priori consistent} map network, the choice of the base shape should not
affect the results. In practice, however, functional maps are represented in a reduced basis and are
not perfectly consistent, which can introduce strong bias in the subsequent analysis.

\begin{figure}[t!]
  \centering
  \includegraphics[width=.95\linewidth]{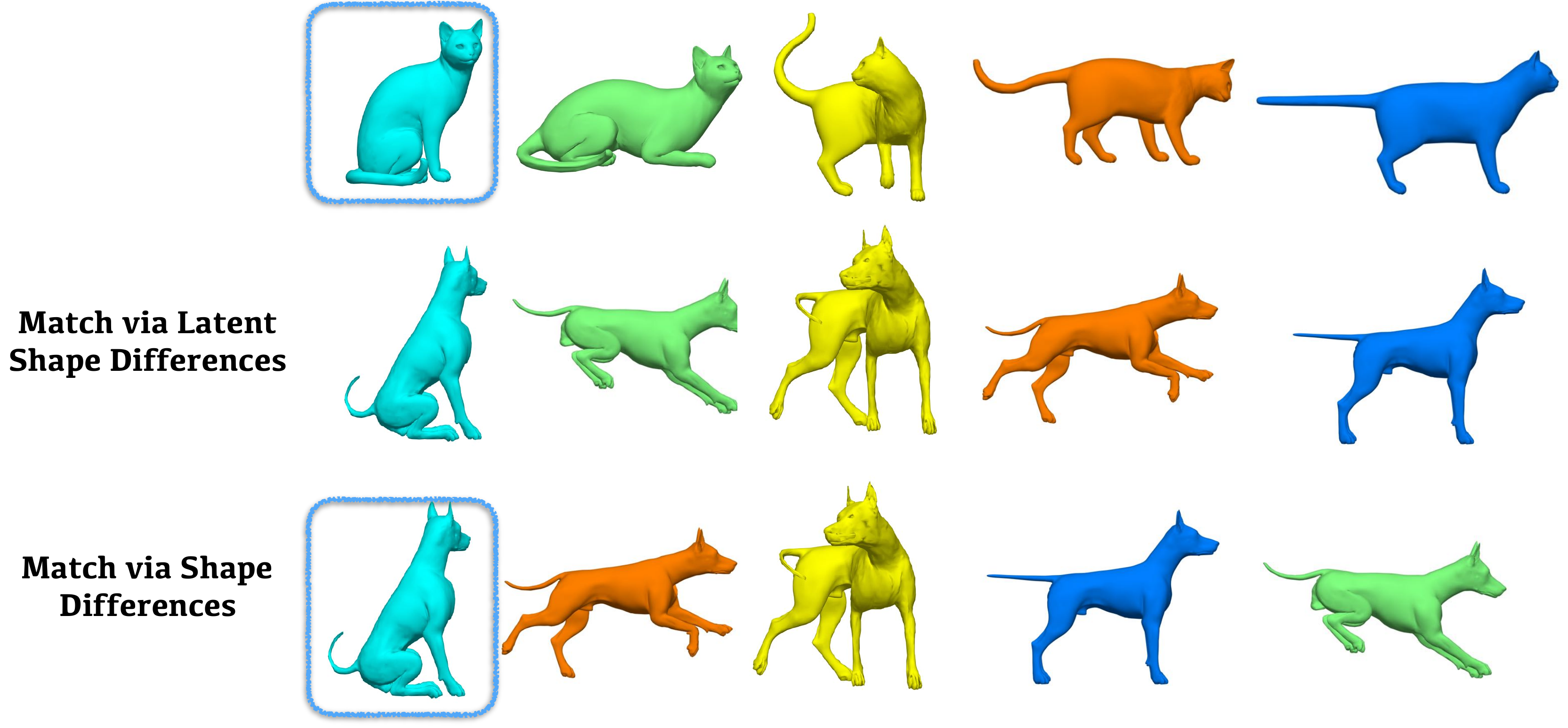}
\vspace{-0.5mm}
  \caption{Simultaneous analogies between a collection of cats and dogs without maps across them. The ground-truth correspondences are indicated by the color-coding. In the middle row, the latent shape differences recover the ground-truth alignment.
On the other hand, the shape differences fail to recover --- one failure examples using the shape differences with the boxed base shape are shown on the bottom. \label{fig:ret_2}
\vspace{-4mm}}
\end{figure}

To illustrate this effect, we aligned a collection of cats and dogs shown in Figure~\ref{fig:ret_2}
\emph{without any maps across them} using the original and the latent space shape differences.
For
the former, we assume that a pair of shapes, e.g., the boxed animals in
Figure~\ref{fig:ret_2}, to be used as bases in each cluster, and computed the
eigenvalues of the respective shape differences as descriptors. 
On the other hand, we used the eigenvalues of the latent shape differences as the descriptor for
each shape in the collection, without any a priori information.
The alignment result based on the above descriptors in shown in the bottom two rows of
Figure~\ref{fig:ret_2}. Note that, when using the approach of~\cite{Rustamov2013} even after fixing
the corresponding base shapes, \emph{none} of the base shape choices led to the correct result. We
demonstrate one such result obtained by fixing the base shapes to be the ones shown in the blue
boxes.  Meanwhile, as shown in the middle row, using the latent shape differences results in the
ground-truth alignment. Note that the same experiment has been conducted in~\cite{Rustamov2013} (see Figure 13 therein), however, to obtain the exact alignment, the authors used \emph{all pairwise} shape differences. 

\begin{figure}[t!]
  \centering
  \includegraphics[width=1\linewidth]{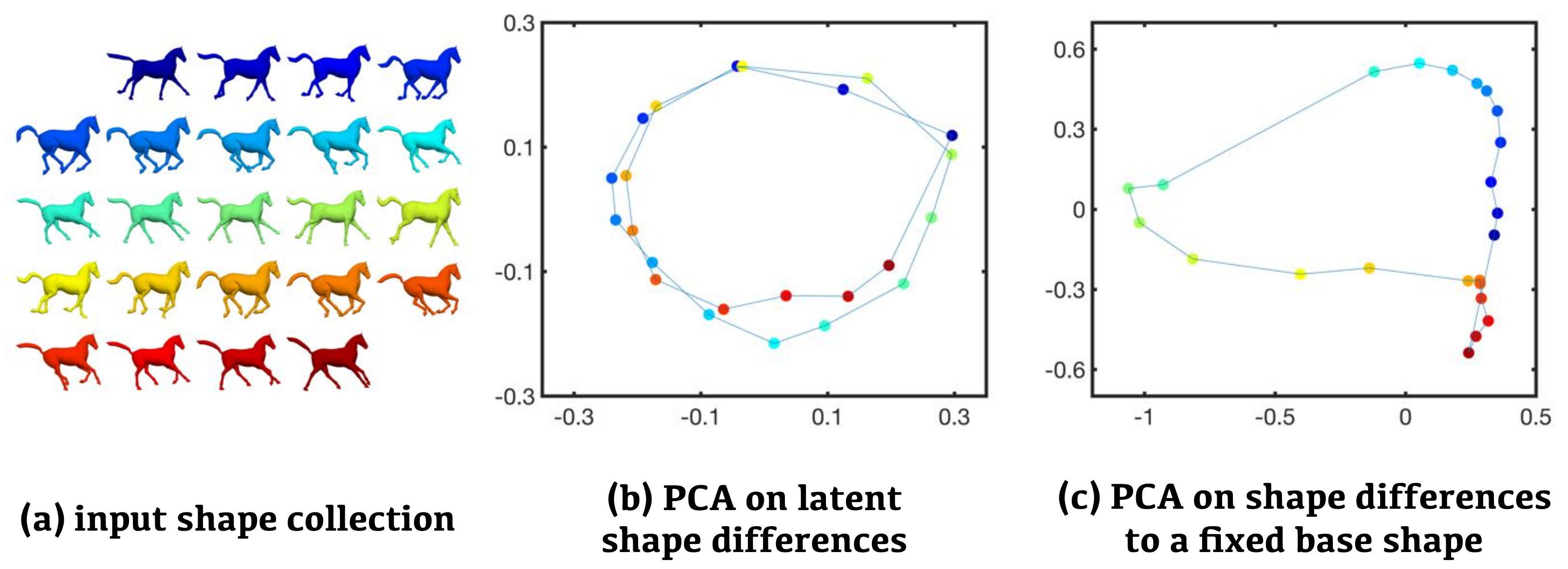}  \caption{\label{fig:layouts} (a) $23$ frames of a galloping horse. With a given FMN of chain topology, we computed the latent and original original shape differences as signatures of the shapes: (b) PCA layout of the latent shape differences; (c) PCA layout of the original shape differences.  
  }
\end{figure}

\myparagraph{Compatibility with sparse map networks:} Another key advantage of our latent
representation is its ability to extract information from sparse map networks. As observed in
previous works \cite{huang2014functional}, functional maps between similar shapes are typically much
easier to compute. On the other hand, establishing functional maps from a fixed base shape to
\emph{all} other shapes in the collection can lead to significant errors.  To illustrate this, we
consider a sequence of $23$ frames of galloping horses shown in Figure~\ref{fig:layouts}(a), and
assume that only functional maps between consecutive frames are given, resulting in a sparse FMN with
chain topology.
Figure~\ref{fig:layouts}(b) demonstrates that, even when extracted from the sparse FMN, the LSSDs
recover the cyclical structure of the collection, while using the shape differences from the
base shape, $S_{12}$, computed by composing the given functional maps, leads to an erroneous
embedding, as shown in Figure~\ref{fig:layouts}(c).

\subsection{Projected Latent Shape Differences}\label{sec:projected_lsd}
Besides encoding each shape in the collection, the latent shape differences also give access to
detailed information about the deformation, including the local changes in different shape regions
in a purely algebraic way.

A link between actual deformations across shapes and functional distortions induced by the
respective shape differences has been established in~\cite{Rustamov2013} -- namely, a function, such
as an indicator function of a region, will be modified by the shape difference, if it is supported
on region undergoing a deformation. It has been further shown in~\cite{Rustamov2013} (see Section 6) that the area-based
(resp. conformal) shape difference is an identity operator if and only if the underlying map is
area-preserving (resp. conformal).

In this section, we propose a novel \emph{projection} operation on the latent shape differences, the
key observation is that we can suppress a functional deformation by modifying the shape difference
so that it acts like an identity operator on certain functional subspace expressing the deformation
of interest, which, in the following allows us to perform partial shape analogies.

Suppose that we are given a set of shapes $\{S_i\}_{i = 1}^n$ and a FMN $\G$, and let
$\{D_i\}_{i = 1}^n$ be the LSSDs computed using Algorithm~\ref{alg:latentdiffs}.
Now we consider a set of functions $F = [\alpha_1, \alpha_2, \cdots, \alpha_k]$, where $\alpha_i's$
are $k$ orthonormal basis functions on the latent shape, i.e., $F^T F = Id_k$.  We construct a
\emph{projected latent shape difference} using $F$ and $D_i$ as follows:
\begin{equation}\label{equ:projection}
	P_i(F) = D_i (Id - F F^T) + F F^T. 
\end{equation}

It is easy to verify that $P_i(F) \alpha = D_i \alpha$ if $\alpha$ is orthogonal to the subspace spanned
by functions in $F$, and $P_i(F) \alpha = \alpha$ if $\alpha$ is spanned by the functions in $F$.
Intuitively, if $F$ contains the full basis on the latent shape, then $P_i(F) = Id$, which forces
latent shape difference to correspond to an area-preserving or conformal map, depending on the type
of $D_i's$.

\section{Shape Collection Comparison}\label{sec:comparison}

Several approaches have been proposed for detecting geometric variability that exists within a given
collection of shapes connected by functional maps, e.g., ~\cite{Rustamov2013,Huang2017}. In this
section, we show how our latent-based shape representation can be used for detecting and analyzing
and differences \emph{across} different shape collections, or two subsets of a larger
collection. Namely, given a set of shapes $\mathcal{S}$, a FMN $\G$ and a partition
$\mathcal{S} = \mathcal{S}^A \bigcup \mathcal{S}^B$, we aim to capture the difference between
$\mathcal{S}^A$ and $\mathcal{S}^B$, while not being sensitive to the global variability that exists
within $S$. This problem arises especially when trying to detect the detailed geometric properties
that are responsible for the differences between shape classes (e.g., healthy vs unhealthy organs),
while factoring out the ``normal'' or ``common'' variability within the collection.

\paragraph*{Global variability}
Before approaching this problem, we first propose an algebraic approach for detecting global
variability within a collection. Our observation is that, in light of
Section~\ref{sec:projected_lsd}, suppressing global variability should lead to projected LSSDs that
are indistinguishable from each other. Namely, we would like to find a basis $F$ such that the
latent shape differences projected onto $F$ are as close as possible. For this, we first introduce a
term that measures the difference of the norms between the original and projected latent shape
differences:
\begin{equation}
	\delta(S_i, S_j, F) = \Vert D_i - D_j \Vert_{\mbox{Fro}}^2 -  \Vert P_i(F) - P_j(F) \Vert_{\mbox{Fro}}^2
\end{equation}

According to the following lemma, the change is always non-negative and can be written in a quadratic form. 
\begin{lemma}\label{lem:delta_energy}
	If $F^T F = Id$, then 
		\[\delta(S_i, S_j, F) = \mbox{Trace}(F^T (D_i - D_j)^2 F) \geq 0 .\]
\end{lemma}

It is natural to optimize for a function $\alpha_{\mbox{intra}}^*$, which maximizes the global change of distances within the collection, i.e., 
\begin{equation}\label{equ:intra}
	\alpha_{\mbox{global}}^* = \mbox{arg}\max_{\alpha} \sum_{i, j} \delta(S_i, S_j, F), s.t. \alpha^T \alpha = 1.
\end{equation}
In other words, after suppressing the functional deformation related to $\alpha^*$, the shapes are
\emph{maximally} brought together.  According to Lemma~\ref{lem:delta_energy}, $\alpha^*$ is given by the
eigenfunction associated with the largest eigenvalues of $\displaystyle\sum_{i, j} (D_i - D_j)^2$.

\paragraph*{Cross-collection variability}
Following the same idea above, we formulate the cross-collection variability to be such that after
suppressing it, the clusters $\mathcal{S}^A$
and $\mathcal{S}^B$
should become closer to each other, while maintaining their inner structure. In other words, we aim
to simultaneously maximize the changes of distances across shapes in different clusters, and
minimize those within the same cluster.

Putting these two goals together, we construct:
\begin{align}\label{equ:inter}
\begin{split}
\alpha_{\mbox{cross-collection}}^* = &\mbox{arg}\max_{\alpha^T \alpha = 1} \sum_{(i, j)\mbox{ across clusters}} \delta(S_i, S_j, F) - \\
&\sum_{(i, j)\mbox{ within same cluster}} \delta(S_i, S_j, F).
\end{split}
\end{align}

\begin{figure}[t]
  \centering
  \includegraphics[width=0.9\linewidth]{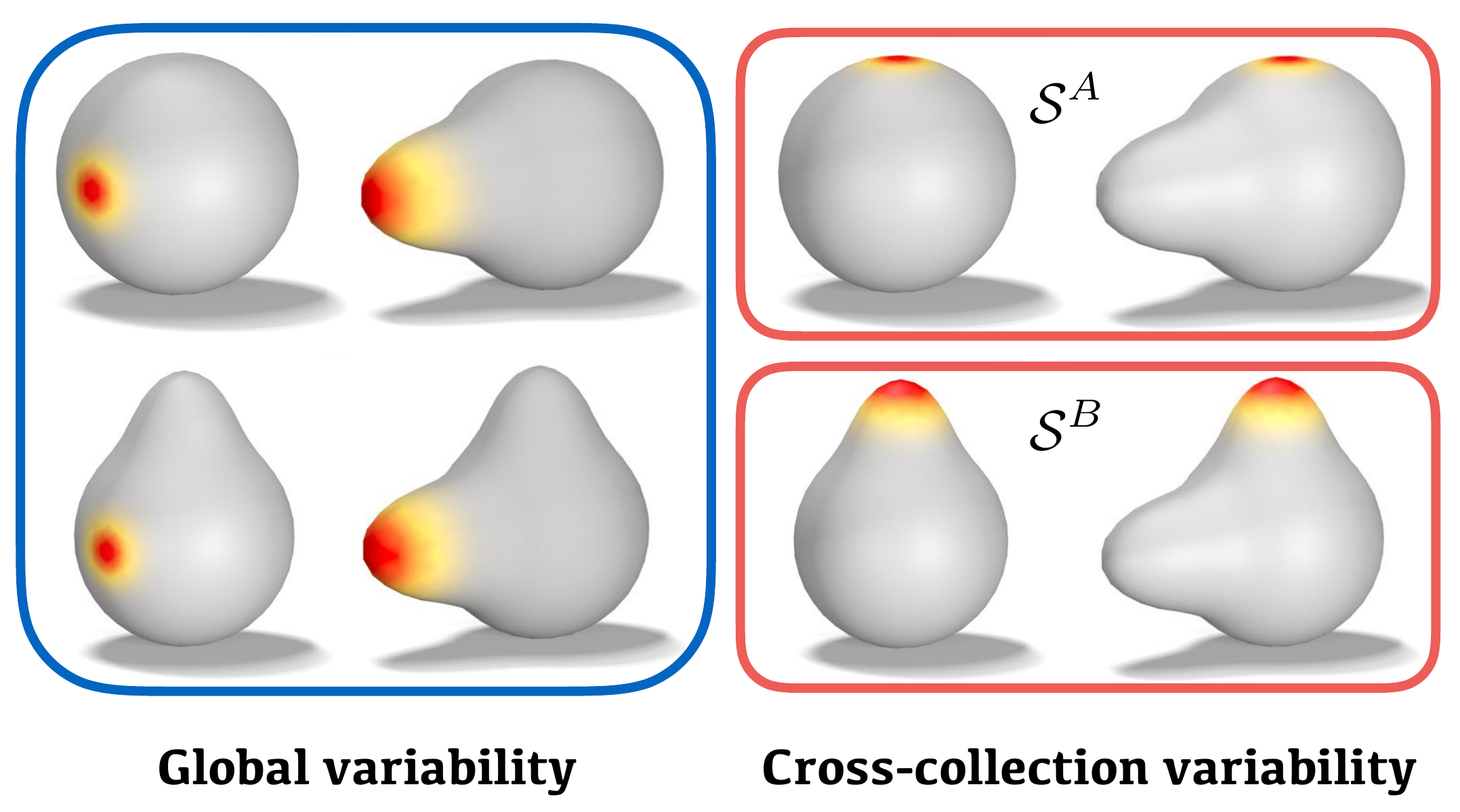}
  \caption{\label{fig:illustration}
  			The global variability of four deformed spheres (in the blue box) and the cross-collection variability regarding the partition $\mathcal{S}^A$ and $\mathcal{S}^B$ (in the red boxes) detected by our algorithms. Note that the horizontal bump (global variability) is of twice the magnitude of the vertical one (cross-collection variability).}
\end{figure}

As an illustration, in Figure~\ref{fig:illustration}, we demonstrate the optimizers
$\alpha_{\mbox{intra}}^*, \alpha_{\mbox{inter}}^*$ respectively.  Since the horizontal bump is of
twice the size of the vertical one, to maximally reduce the intra-variability, one should suppress
the horizontal deformation.  Meanwhile, it is intuitive that cluster $A$ and $B$ are distinguished
by magnitudes of the vertical bumps, which should be detected as cross-collection variability.

Finally, we point out that, though not being equivalent, there is a connection between the
formulations above and the one for detecting global variability proposed in~\cite{Huang2017}. In
fact, we can use results from both approaches for cross-validation. We refer interested readers to
Appendix for the statement and proof of this connection.

\section{Applications in Learning}
\label{sec:learning}

On the (deep) learning side of our exposition we study how our representation of the latent shape difference operators, can be used 
as the input that neural networks will rely upon to reason about 3D data. A key property of this input representation is that it is encoded as a small size matrix -- i.e. it provides a regular structure amenable to convolutions.  CNNs rely and take advantage of the spatial proximities found in regular-grid data such as images. Analogously, according to our formulation for LSSDs, we have, e.g. in Eq.~\ref{equ:area_latent}, $D_i^A(k, l) = y_k^T y_l$, where $y_k$ is the $k$-th latent basis function. Thus, instead of spatial proximity, the neighboring entries in our matrix representation encode and provide interactions of function pairs that are close in the ``spectral'' domain.

The first task at which we test the effectiveness of our approach is {\em shape regression} -- here we compare neural-networks that learn to estimate hidden parameters that control the body variation of human-form meshes. Assuming a set of training shapes, with known parameters that are represented as real-valued vectors, our networks learn to regress the underlying parameters for new unseen shapes.

The second task we explore is that of 3D point-cloud {\em reconstruction}. Quite differently from the regression task, here we test how a novel network 
that inputs a latent difference matrix can learn to reconstruct a 3D point-cloud version of the underlying mesh. This problem is closely related to shape reconstruction from intrinsic operators, which was recently considered in \cite{boscaini2015shape,Corman2017} where several advanced, purely geometric, optimization techniques have been proposed that give satisfactory results in the presence of full information \cite{boscaini2015shape} or under strong (extrinsic) regularization \cite{Corman2017} --- but they also demonstrate the many challenges posed by this type of reconstruction. In contrast, we show that by using the context of a collection and learning machinery, real shapes can be recovered rather well from their latent difference operators, and moreover that entirely new shapes can be synthesized using the algebraic structure of difference operators.

One possible concern with our approach is that it requires an initial functional map network, which can potentially restrict the amount of training data available. However, as we show in Section~\ref{sec:experimental_result} even for collections of moderate size, consisting of a hundred to two hundred shapes, our networks are sufficiently regularized and allow for very powerful and effective learning.

\subsection{Localized Latent Shape Difference}\label{sec:llsd}
Localized shape deformation is a useful tool for shape analysis and synthesis in geometry processing. The algebraic form of our latent representation makes it easy to manipulate, meanwhile, the geometric information encoded in it allow us to access the local geometric features. 

Given shapes $A$ and $B$ with their respective LSSDs $D_A$ and $D_B$, and a set of  basis functions $F$, expressed in the basis of the latent shape and which is supported on localized region $S$ on the shapes, we can construct an operator that acts as $D_A$ on $S$ and as $D_B$ on the complement of $S$ as follows: 
$D^{A,B}_{\text{part}} = D_A (Id - F F^T) + D_B F F^T.$ Note that this expression resembles Eq. (\ref{equ:projection}) above, but where we consider $D_B = Id$, so that one of the interpolated shapes is the latent shape itself. Using $D_{\text{part}}$ allows us to construct shapes by mixing different parts or regions of existing shapes, leading to localized interpolation/shape analogy, as we will show in Section~\ref{sec:recon}.

%!TEX root = main.tex
\section{Main Experimental Results}\label{sec:experimental_result}

\subsection{Applications in Learning}
In this set of experiments we explored how latent shape differences can be used within the context of a 3D deep learning pipeline. As mentioned in Section \ref{sec:learning} our latent differences provide a new representation of geometry with unique characteristics, suggesting its use in 3D-ML applications.

\subsection{Data-generation}
\begin{figure}[ht!]
  \centering
  \includegraphics[width=80mm, scale=0.6]{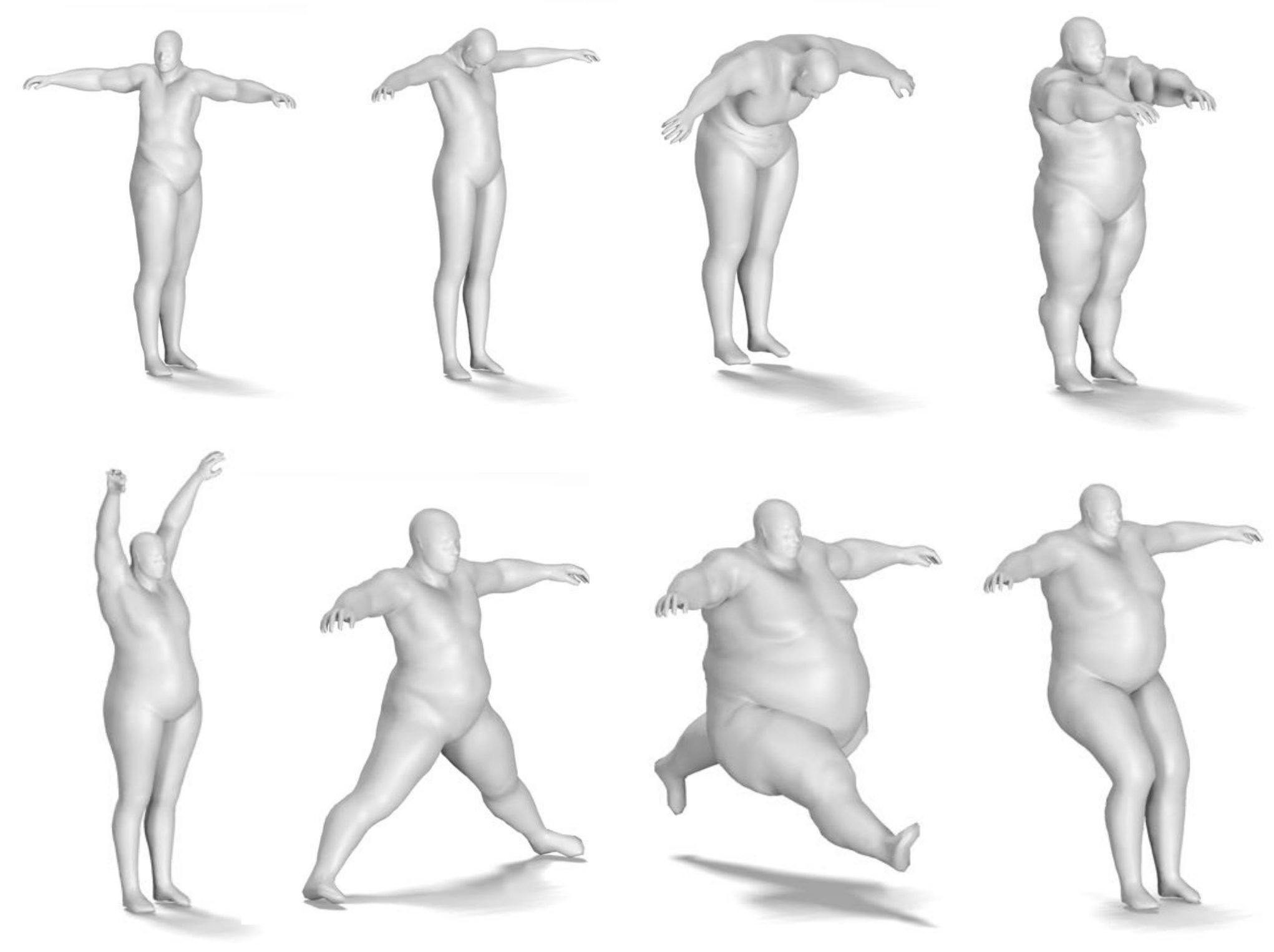}
        \caption{\label{fig:scape_shapes} Example synthetically generated meshes used within the leaning-based pipelines of Section~\ref{sec:experiments}, displaying a randomly selected mesh of each pose-class.}
\end{figure}

For the experiments of Sections~\ref{sec:learn-regression} and \ref{sec:recon} we generated $400$ human shape bodies in eight different poses using the open-source implementation \cite{Deep3DPose} of the SCAPE method \cite{scape}. In \cite{Deep3DPose}, body variations are controlled with $12$ latent parameters $\in [0, 1]$, which informally encode shape attributes such as height, leg-girth, belly protrusion, etc. To generate our shapes we sampled uniformly i.i.d.~each of the aforementioned parameters and considered eight modifications of the standard T-pose.
See Figure~\ref{fig:scape_shapes} for a sample of the resulting meshes. It is worth noting that the produced meshes share the same combinatorial tessellation on 6,449 vertices, which facilitated the construction of pairwise functional maps in this collection. For the following experiments of regression and reconstruction, we used a train-test-val split with $75\%, 15\%, 10\%$ of this dataset respectively.

\subsection{Regression}
\label{sec:learn-regression}
In this experiment, we assess the efficacy of a neural-network in regressing the body-generating parameters under different types of input representations. Concretely, we compare the responses between two types of input: point-clouds with $1024$ points sampled uniformly area-wise from each mesh and area-based latent differences. We explore the effect of several design choices in the construction of our differences.
First, we consider different topologies of the underlying Functional Map Network (FMN). These include the complete graph but also much sparser versions based on the $k$-nearest-neighbors ($k \in \{10,20\}$) of each shape. Second, we vary the dimensions of the latent {\em bases} which crucially effects the size of the difference matrices. We use the $50$ LBO eigenvectors with the smallest eigenvalues to express all functional maps and the Euclidean norm of these spectra to define a distance for the construction of the $k$-nearest-neighbors. Last, we train our neural-networks to minimize the Mean-Square-Error (MSE) between their predicted and the ground-truth shape generating parameters. Note, that since these parameters are independent of a shape's pose, pose-variations of this dataset act as ``nuisance'' variables that the networks have to explain-away. 

\subsubsection{Comparing architectures: protocol}
To select a good point-cloud (PC) based architecture we evaluate three PointNet-like networks \cite{qi2016_pointnet} that use encoding/decoding schemes like those of \cite{achlioptas2017latent_pc}. These architectures have shown excellent results in tasks involving 3D point clouds, including classification, part-segmentation and generation and provide a strong baseline. Concretely, our point-based architectures have three layers of convolutional encoders, followed by a feature-wise max-pool and either two or three layers of FC-ReLUs that act as decoders. To strengthen our comparisons, we calibrate each PC architecture to have a distinct number of training parameters and train it with several learning rates to obtain from a pool of $15$ models, the one with the best performance (see Appendix Sec.~\ref{sec:PC-archs} for more details).

At the same time, we consider two types of architectures when the input is a latent-difference: Multi-Linear Perceptrons (MLPs) and Convolutional Neural Networks (CNNs). Across all experiments, these MLPs are four layer deep and the CNNs have two layers of convolutions leading to a third FC layer (see Appendix Sec. ~\ref{sec:recon-arch} for more details).

\begin{figure}[ht]
  \centering
  \includegraphics[width=70mm, scale=0.7]{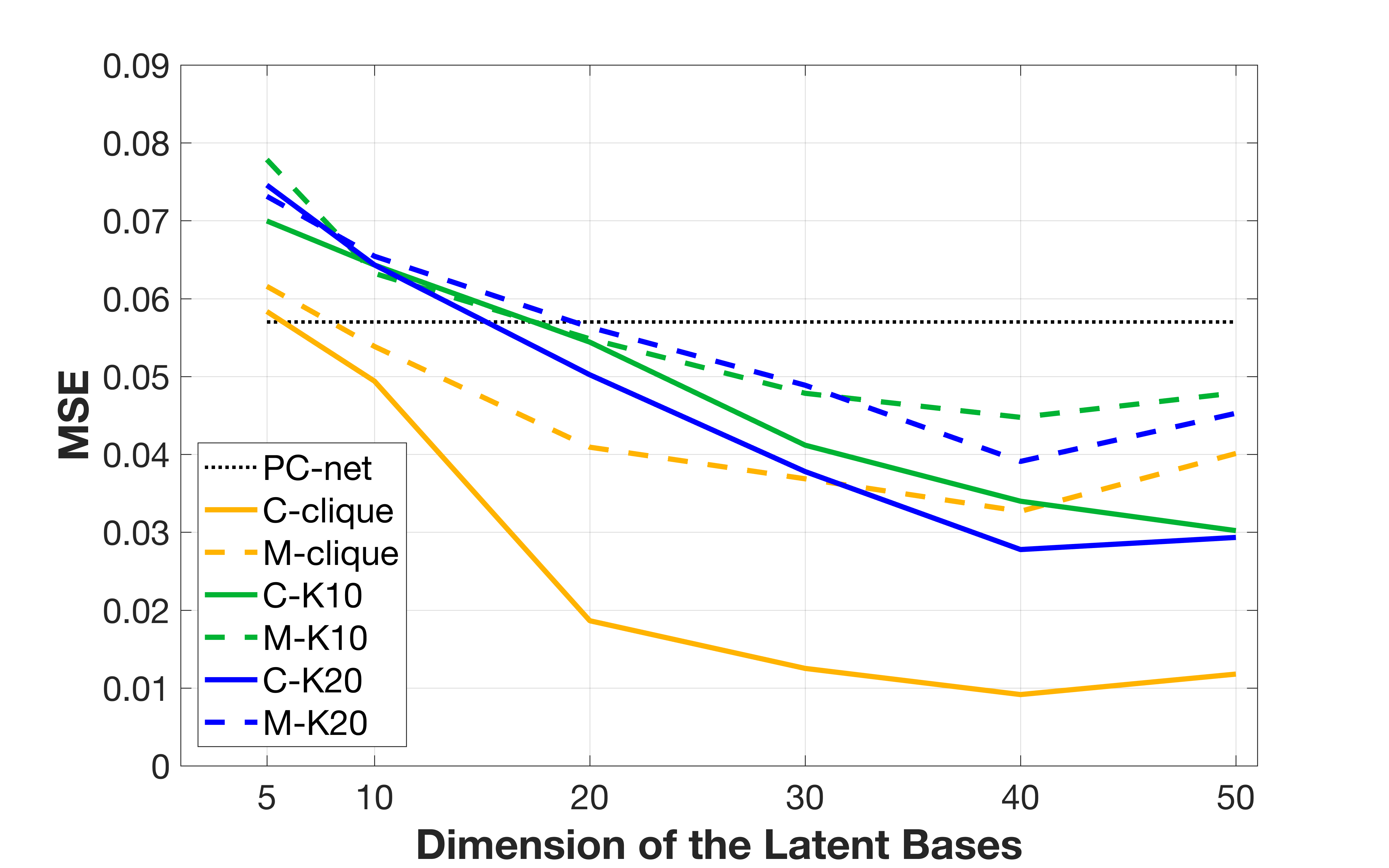}
        \caption{\label{fig:mse} Regression-based comparison of different input modalities and neural-nets. The $y$-axis depicts the Mean-Square-Error (MSE) on the {\em test} split. The $x$-axis corresponds to the number of consistent functions each shape was associated with. The dashed line is the point-based baseline (best of $15$ such models). Models starting with M (solid lines) are MLPs and starting with C (scattered points) CNNs. K10/K20: sparse 10/20-nearest-neighbor FMN topologies. Clique stands for the clique FMN topology. The results are averages of 5 random seeds.}
\end{figure}

Figure~\ref{fig:mse} shows the MSE between the predicted vectors and the ground-truth for the test shapes in a variety of conditions. 
The reported MSE is the average over five random data-splits and weight initializations of the neural nets. The networks are trained maximally for $500$ epochs and the displayed MSE correspond to the model (epoch) that optimized the validation split. The dashed-line shows the performance of the best over-all point-based architectures.

\paragraph{Discussion.}
\label{parag:mse-discusion}
Figure~\ref{fig:mse} reveals several trends. First, shape difference CNNs perform better than MLPs and both perform significantly better than point-based nets for a wide variety of different configurations. Second, there seems to exist a sweet-spot in the range of 35 and 45 latent bases --- which consistently produces better results across different network topologies. Third, denser topologies give rise to better results with the clique FMN achieving the best performance. In Table~\ref{table:mse_gen_error} we include some complementary information to that of Figure~\ref{fig:mse}. 
Its first row contains the MSE measurements for the point-base network (PC column) and some MLP/CNNs configurations (clique or 20-nearest-neighbors topology with $40$ basis functions). The second row reports the generalization error (difference between test and training MSE) of each architecture. The architectures seem to over-fit in a similar fashion percentage-wise, but crucially the difference-based ones, do so at significantly lower values. The last two rows report the MSE and the average $L_1$ distance between the predictions and ground-truth when we train these networks for 1,000 instead of 500 epochs.

\begin{table}[ht]
    \begin{center}
    \begin{tabular}{c c c c c}
        \hline
        Metric   & PC & MLP-Clique & CNN-20 & CNN-Clique\\
        \hline
        \hline
        MSE@500  & 0.057 & 0.033 & 0.027 &\bf{0.009}\\
        GE@500   & 0.020 & 0.009 & 0.010 &\bf{0.003}\\
        MSE@1K   & 0.061 & 0.032 & 0.013 &\bf{0.005}\\
        $L_1$@1K & 0.192 & 0.134 & 0.086 &\bf{0.050}\\
        \hline
    \end{tabular}
    \caption{Complementary statistics of Fig.~\ref{fig:mse}, see  \ref{parag:mse-discusion} for details. GE stands for Generalization Error. For reference, if our output prediction was the average of the training examples, then the average $L_1$ would be 0.245 and the MSE 0.08.}
    \label{table:mse_gen_error}
    \end{center}
\end{table}

\subsection{Reconstruction}\label{sec:recon}
In the second set of deep learning experiments we demonstrate how we can reconstruct a point-cloud derived from a 3D mesh based on the corresponding latent area-based difference operators. To achieve this we use a wider and deeper version of our previous CNN with the regression-optimal input: difference matrices of dimensions $40 \times 40$, based on a clique FMN. The new network is comprised of 5 layers, with the first two layers being convolutional and the remaining three FCs (see Appendix Sec.~\ref{sec:recon-arch} for more details). The output of this network is $4,096\times3$ real-numbers which are trained to have minimal Chamfer-(pseudo)-distance, from the corresponding ground-truth point-clouds that are comprised also of $4,096$points (similar to \cite{fan2016_a, achlioptas2017latent_pc}).

\begin{figure}[ht!]
  \centering
  \includegraphics[width=70mm, scale=0.7]{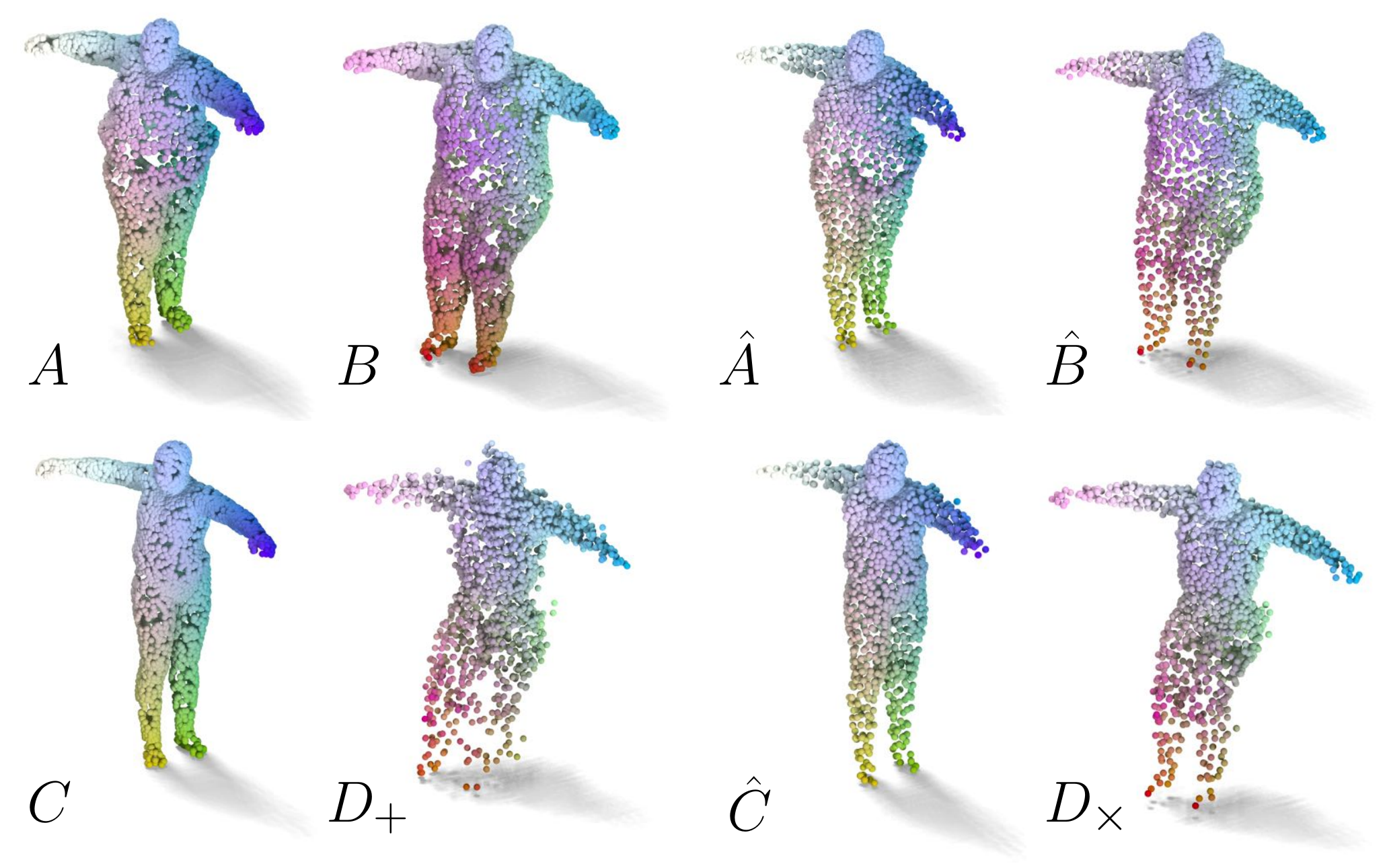}
        \caption{\label{fig:analogy_1} Reconstructed point-clouds from area-based latent differences. Point-clouds $A, B, C$ belong to the ground-truth {\em test} split and  $\hat{A}, \hat{B}, \hat{C}$ are their corresponding reconstructions. $D_{\times}$ and $D_{+}$ complete the analogy of a shape that {\em is to C what B is to A}. Reconstruction $D_{+}$ is based on traditional vector code arithmetic, while $D_{\times}$ is the one based on shape difference operator algebra.}
\end{figure}

Figures~\ref{fig:analogy_1} and \ref{fig:analogy_2} demonstrate the quality of the learned reconstructions along with the capacity of our representation for doing semantically-rich shape synthesis operations, such that of constructing new shapes (not present in the original shape collection) based on shape-analogies. First, to visually inspect the reconstruction quality compare the {\em ground-truth} point-clouds: $A, B, C$ with their corresponding reconstructions $\hat{A}, \hat{B}, \hat{C}$. These ground-truth point-clouds belong in the test split and their reconstructions have successfully captured both the underlying pose and the body structure. While this is true, it is also evident that some high-frequency geometric information (mostly around the hands) has not been recovered. Despite these artifacts, these results are remarkable, given previous attempts at shape reconstruction from difference operators \cite{boscaini2015shape,Corman2017}, which only work by combining both area and conformal differences and work in very restricted settings under strong regularization. 

We also test the generalization power of the network by synthesizing shapes $D_{+}$ and $D_{\times}$ that try to have a similar pose-wise and body-wise relation to the point-cloud $C$, as the relation that point-cloud $A$ has to $B$ (to form an analogy).  To construct $D_{+}$ we decode (i.e. reconstruct) the neural-network's latent code corresponding to the additive formula: $l_C + l_B - l_A$, where $l_K$ is the output activations of the first FC layer when the input is shape $K$. This is the traditional practice in performing analogies with the latent-codes of a deep-net \cite{DBLP:journals/corr/MikolovSCCD13,3dgan, achlioptas2017latent_pc}, based on latent vector arithmetic. In a different way that better reflects the nature of difference operators, we also reconstruct the result of the multiplicative formula $D_B \times D_A^{-1} \times D_C$ with $D_X$ being the difference operator of shape $X$. Here we directly exploit the matrix nature of our representation which enables this type of algebra. The result of this approach is $D_{\times}$. It is interesting to observe that this reconstruction, ($D_{\times}$), results not only in less noisy point-clouds compared to $D_{+}$; but also in semantically more appropriate structures, e.g. in Fig~\ref{fig:analogy_1}, $D_{+}$ reflects less prominently the expected sitting pose and in Fig~\ref{fig:analogy_2}, $D_{+}$ has more muscular arms than expected.

\begin{figure}[ht!]
  \centering
  \includegraphics[width=65mm, scale=0.6]{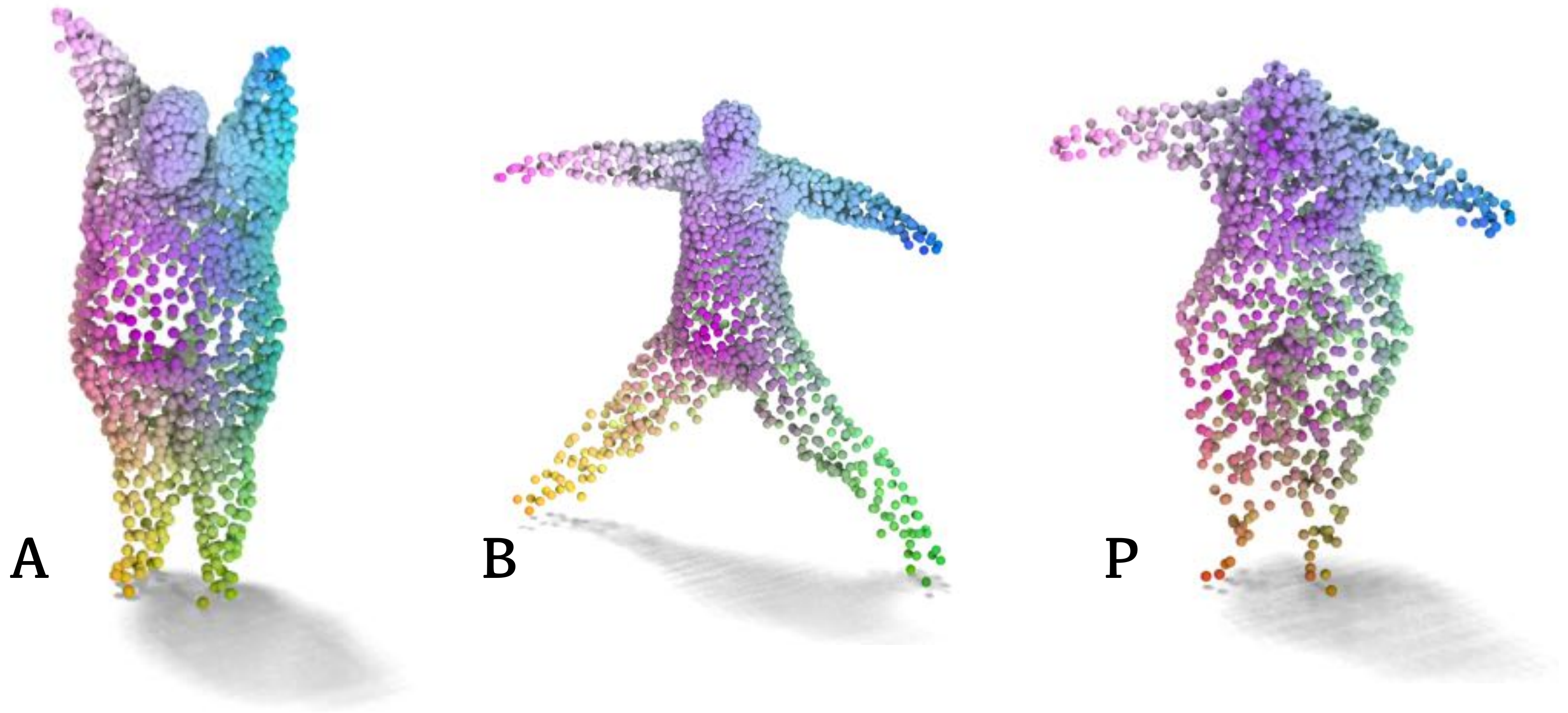}
        \caption{\label{fig:interpolation1} Synthesis of a shape $P$ that is similar to $A$ in the lower body, while being similar to $B$ in the upper body. }
\end{figure}

\begin{figure}[ht]
  \centering
  \includegraphics[width=80mm, scale=1.1]{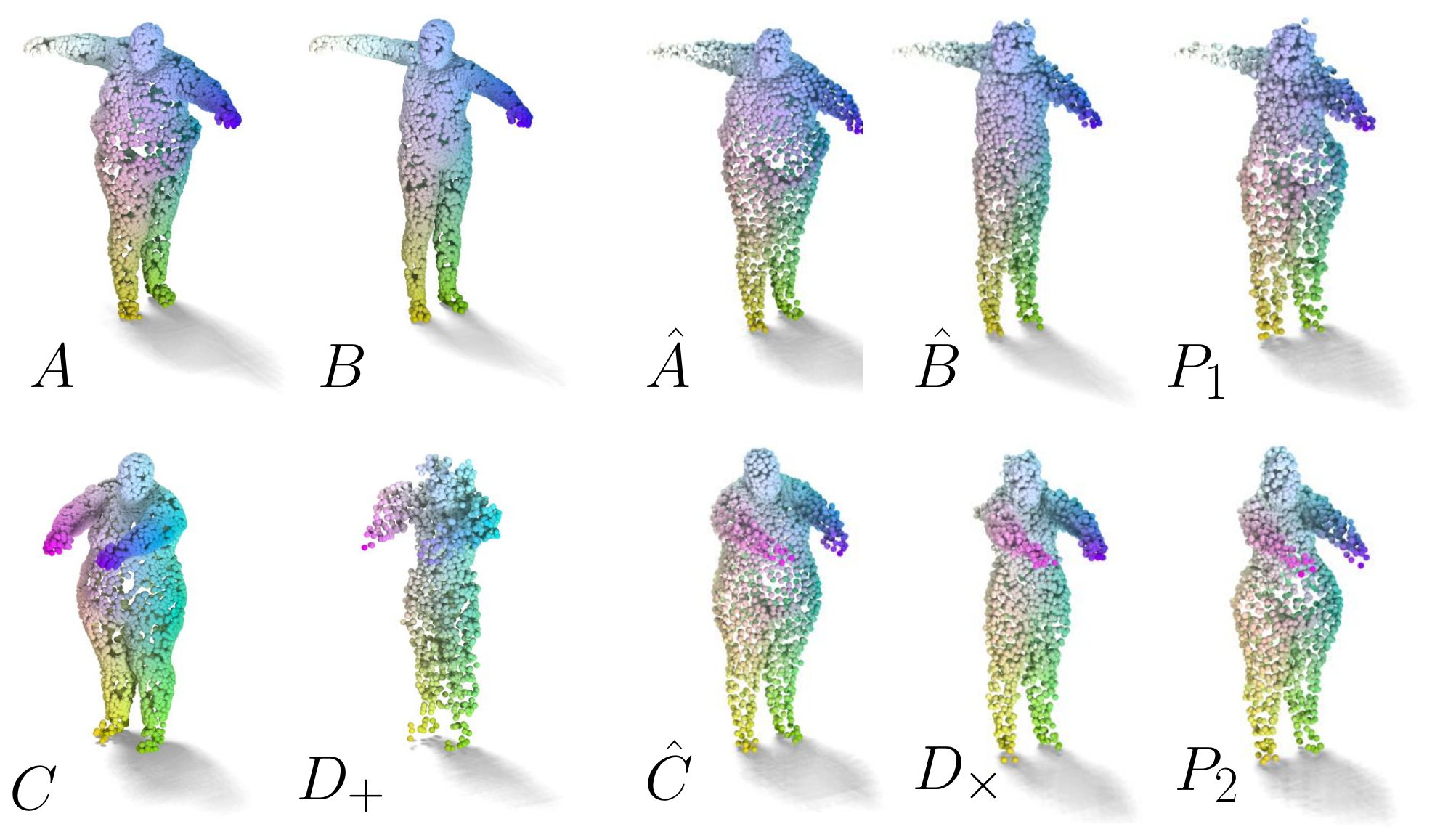}
        \caption{\label{fig:analogy_2} Global and partial shape analogies. Note that $P_1$ has a mixed body type of $\hat{A}$ and $\hat{B}$, and so has $P_2$ but in a different pose following $\hat{C}$ and $D_{\times}$.  }
\end{figure}

\myparagraph{Partial shape analogies}
Moreover, we propose to construct \emph{partial} shape analogies. 
We follow the formulation described in Section~\ref{sec:llsd} -- in parallel to $D_{\mbox{part}}^{A, B}$, we construct $D_{\mbox{part}}^{C, D_{\times}}$ for localized deformation transfer between $C$ and $D_{\times}$. 
We first show in Figure~\ref{fig:interpolation1} a partial body transfer. 
Given the LSSDs regarding $A$ and $B$, we restricted the region of interest to their upper body, and synthesized the LSSD. 
The reconstructed point clouds are shown in Figure~\ref{fig:interpolation1}. Note that $P$ is similar to $A$ in the lower body, while being similar to $B$ in the upper body. 

In Figure~\ref{fig:analogy_2}, we show both the global shape analogies and the partial ones. 
$P_1, P_2$ are the reconstruction result of $D_{\mbox{part}}^{A, B}$ and $D_{\mbox{part}}^{C, D_\times}$, respectively.

\myparagraph{Generalization with computed functional maps}
 
Though the input LSSDs of our network are precomputed with respect to all the shapes in consideration, as mentioned at the end of Section~\ref{sec:latentShape}, we can assign the latent basis to new unseen shapes without recomputing the latent basis.  In particular, we generated a set of new human shape bodies, and for each shape, we searched for its nearest neighbor in the existing collection. We then the kernel matching algorithm \cite{kernel17} to compute an initial map between the new shape and its neighbor in the collection, allowing us to compute its latent representations, as described at the end of Section \ref{sec:latentShape}.

\begin{figure}[ht]
  \centering
  \includegraphics[width=0.8\linewidth]{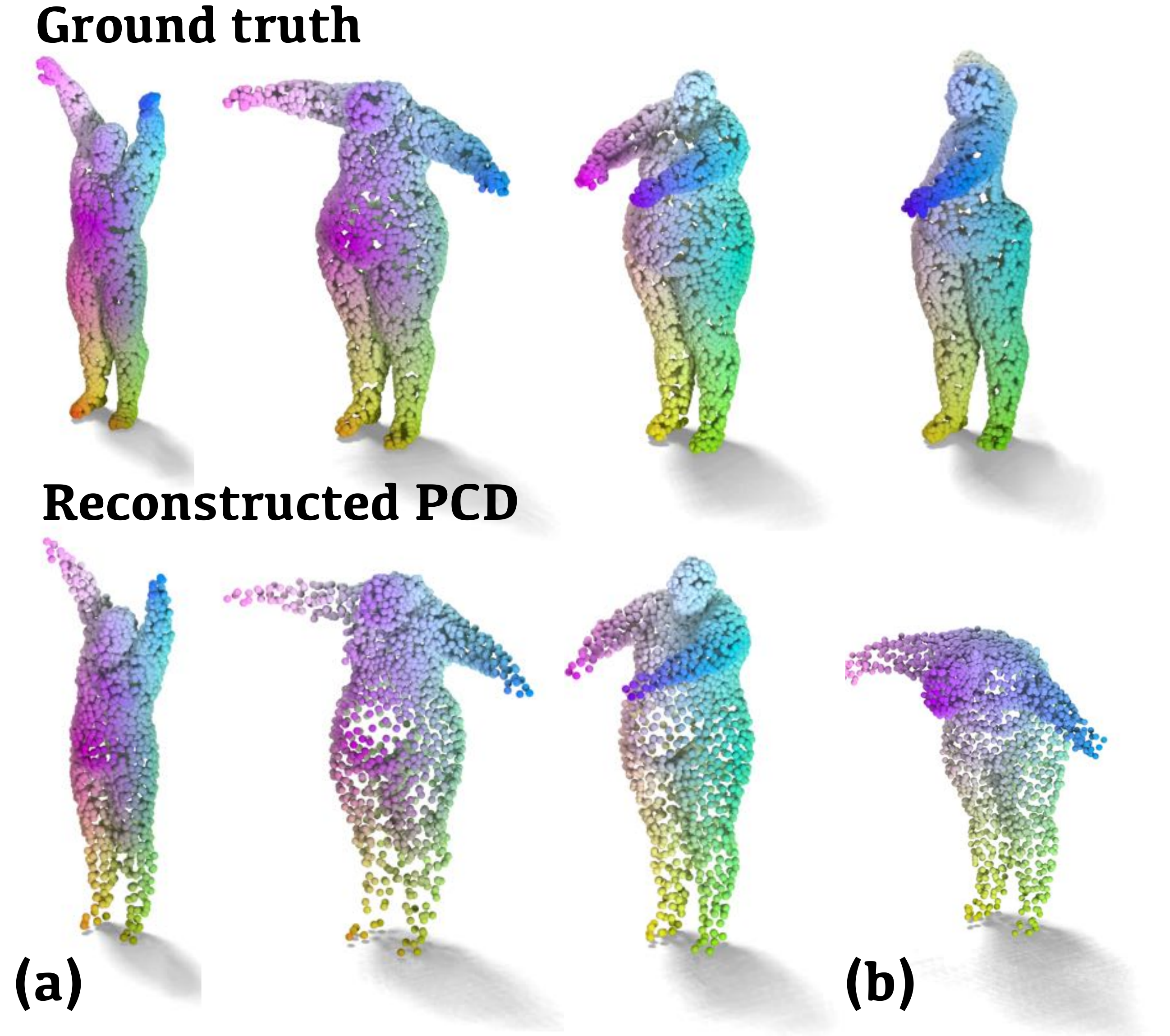}
        \caption{\label{fig:recon_comp} Reconstruction result using computed functional maps on unseen shapes. Top row: ground truth; Bottom row: reconstructed point clouds; (a) $3$ successful cases; (b) a failure case. }
\end{figure}

We show some of the reconstruction results in Figure~\ref{fig:recon_comp}. In particular, we show in~\ref{fig:recon_comp}(b) a failure case.  It is worth noting that though the result has a wrong pose compared to the ground truth, the body type is recovered.  This is due to the fact that in this dataset, the variability across different body types is more prominent, while the number of poses is limited.  Thus the network is expected to put more weight on the features regarding the former, resulting some mismatch poses. We also emphasize that lifting the need of a base shape is crucial in this case, since estimating functional maps across distant shapes is error-prone. Contrastingly, our formulation simplifies such matching procedure. 

\myparagraph{Latent shape interpolation}
We also considered another dataset for the reconstruction task -- the Dynamic FAUST dataset~\cite{dfaust}, from which we sampled $2560$ shapes for training, validation and test. 
For the computational efficiency, we computed the canonical latent basis among a subset of $400$ shapes, and push the basis to the rest $2160$ shapes in the same way above, but using the ground-truth functional maps. 
It is worth noting the shapes in this dataset manifest high extrinsic variability while being near isometry within the poses corresponding to the same character. 
On the other hand, our representation is purely intrinsic, making it challenging to learn features that differentiate the extrinsic change. 

\begin{figure*}[ht]
  \centering
  \includegraphics[width=0.9\linewidth]{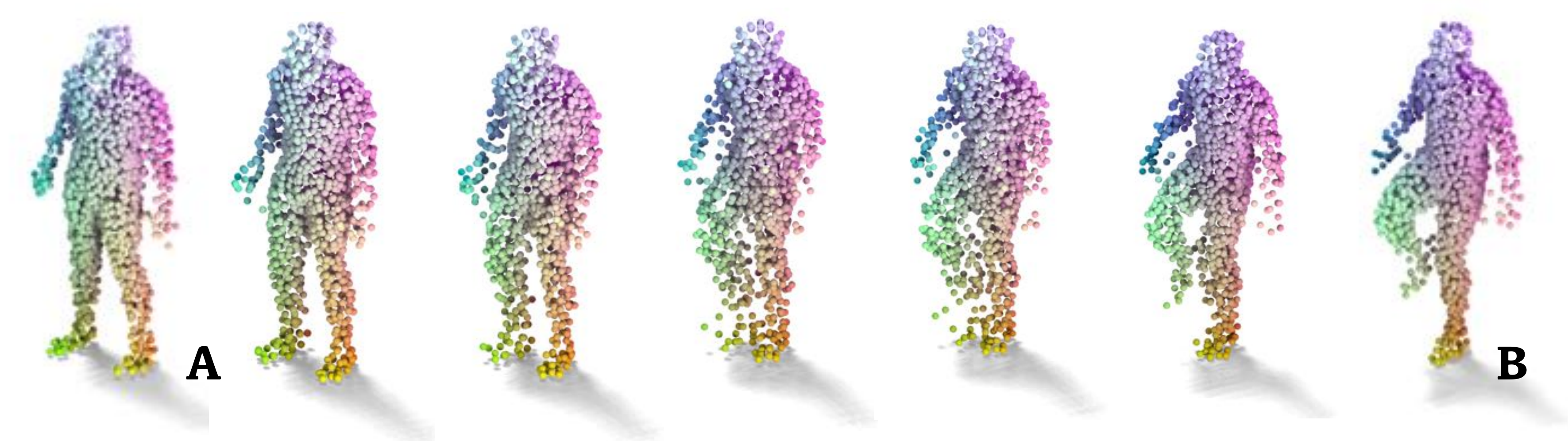}
        \caption{\label{fig:interpolation_dfaust} Latent space interpolation: given two shapes $A, B$, we synthesized new LSSDs by constructing $D_t = (1-t)D_A + tD_B$. The reconstructed point clouds of the $D_t's$ present a continuous deformation from $A$ to $B$. }
\end{figure*}

Here, we demonstrate the advantage of the algebraic form of our representation. 
We selected a pair of shapes $A$ and $B$ from the test set, and construct a sequence of linear interpolation between their latent representations $D_A, D_B$, i.e., $D_t = (1-t)D_A + tD_B, 0<t<1$. 
The output of the network, given $\{D_t\}$, presents a continuous change regarding the knees and upper body. 
In Figure~\ref{fig:interpolation_dfaust}, we selected $5$ output point clouds with respect to increasing $t$ (from left to right), which show a process of raising the right knee and leaning to left from $A$ to $B$.

%!TEX root = main.tex
\label{sec:experiments}

\subsection{Geometric Exploration of Shape Collections}

\begin{figure*}[t!]
  \centering
  \includegraphics[width=0.8\linewidth]{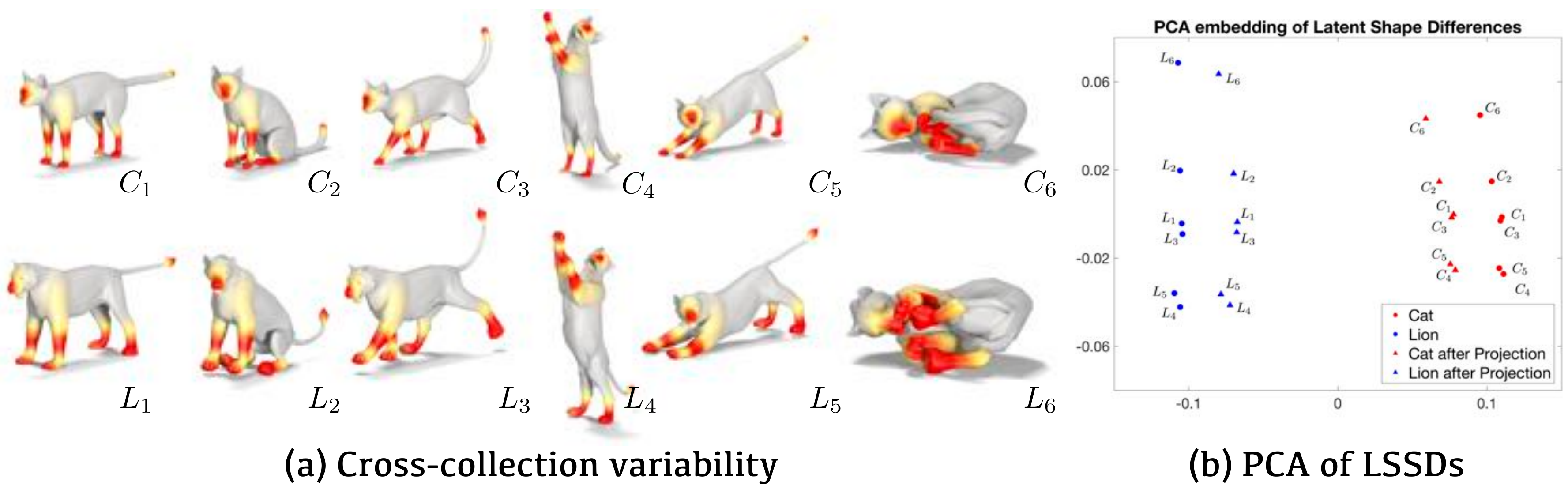}
\vspace{-2mm}
  \caption{\label{fig:two_layer}
Cross-collection variability between a set of cats and lions, as detected by our algorithm.
Note the four paws and the tips of the tails are highlighted, distinctive to each class, despite the presence of the global variability in each collection due to the various poses.
\vspace{-3mm}}
\end{figure*}

\begin{figure*}[t!]
  \centering
  \includegraphics[width=0.9\linewidth]{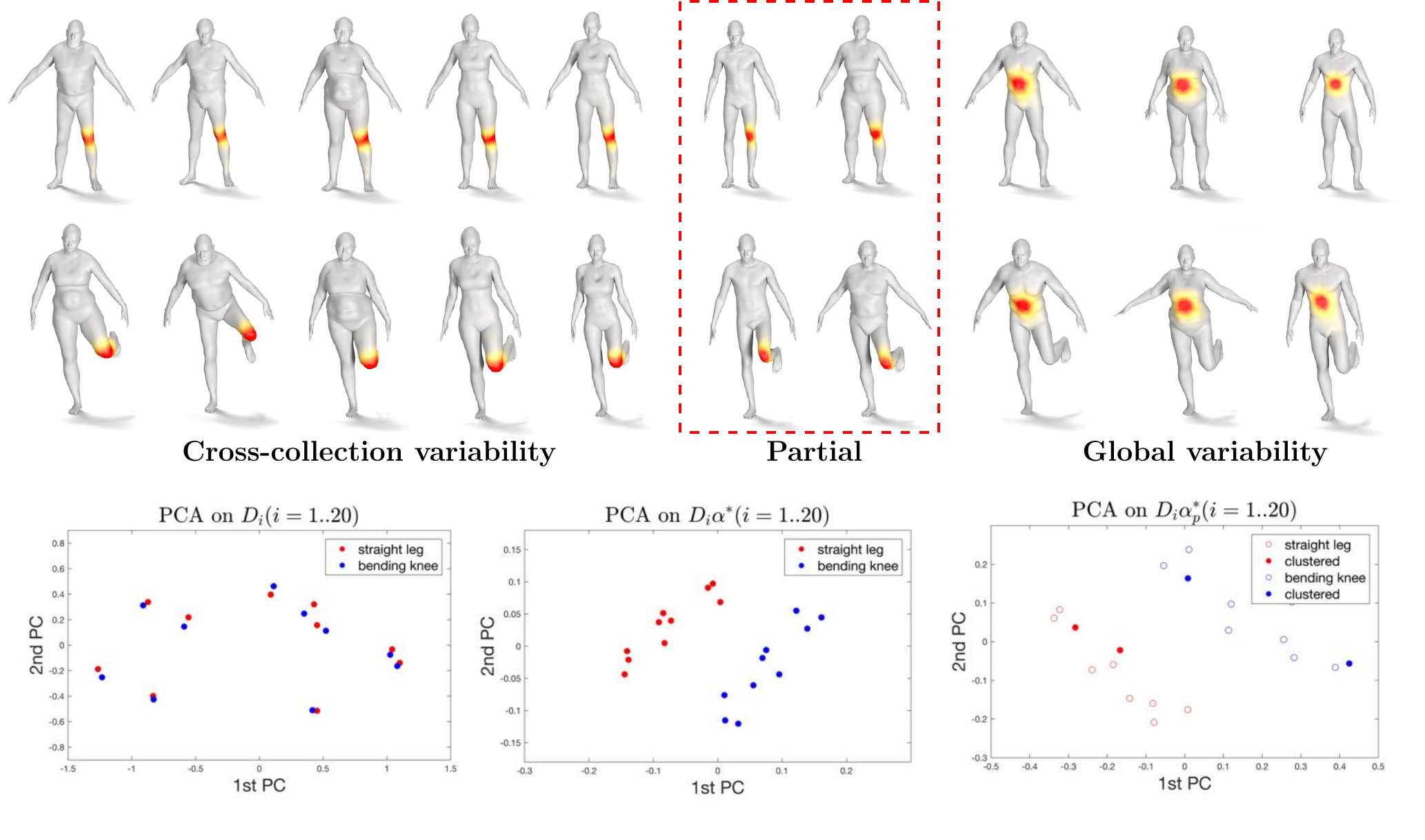}
\vspace{-7mm}
\caption{\label{fig:learnsig} Comparison between two sets of humans, each corresponding to a distinct pose (top
  vs. bottom).  The distinctive functions obtained with our approach capture the change in pose, even with partial
  information, whereas the global variability primarily highlights changes in body type. Moreover, by using the action
  of each latent shape difference on the distinctive function, we can separate the two clusters (shown via PCA in
  bottom row). See text for details.\vspace{-3mm}}
\end{figure*}

In the following experiments we demonstrate the utility of our method for capturing
  cross-collection variability in shape collections, as suggested in Section~\ref{sec:comparison}. In particular, we demonstrate that our method can be applied to real-world data
  (Figure~\ref{fig:bones}) beyond the synthesized shapes, and can be used to compare point clouds
  as well as triangle meshes.  We also demonstrate that our method can extract informative signals
  in a semi-supervised classification task (Figure~\ref{fig:learnsig}). Finally, we demonstrate that
  our method is stable with respect to the input functional maps, as it produces comparable results
  when using \emph{computed} and ground truth maps functional maps (Figure~\ref{fig:two_layer},\ref{fig:bones},\ref{fig:face},\ref{fig:pcd}). 

Throughout the results below, unless stated otherwise, we used area-based LSSDs, which are represented as matrices of size
$60\times 60$ in the reduced basis.  To
construct the FMN, we first compute distances among shapes (using the shape-DNA
descriptors~\cite{Reuter2006}), and form a minimum spanning tree network using these distances. When
considering two clusters, we first form a spanning tree on each, and connect shapes across clusters
using nearest neighbor search.  The methods described in Section~\ref{sec:comparison} optimize for
functions $\alpha^*$ on the latent shape, which we map to functions on the actual shapes in the
collection, resulting in a consistent and informative visualization. 

We applied our method with computed functional maps as input as well. Unless stated otherwise, in the following we used the kernel matching algorithm~\cite{kernel17} for an initial point-wise map, and then converted and refined the maps using functional map techniques (see, e.g., \cite{Functional}).

\paragraph*{\textbf{Heterogeneous Shape Collection Comparision}}
We first demonstrate that our method can capture variability across heteregeneous shape collections,
without relying on point-wise correspondences, through its use of functional maps framework.  For
this, in Figure~\ref{fig:two_layer}, we show the computed distinctive functions highlighting the
difference between a set of cats (each consists of $7207$ vertices) and a set of lions (each
consists of $5000$ vertices), where the cross-collection maps were estimated using the original
functional maps approach \cite{Functional} given a sparse set of landmarks. Note that our method
correctly highlights the snouts, the four paws and the tips of the tails, distinctive to each class,
despite the presence of the global poses variability in the collection.
Moreover, we consider a quantitative validation of the cross-collection variability detected by
  our method, by comparing the PCA embeddings of the LSSDs before and after projection with respect
  to the highlighted functions shown in Figure~\ref{fig:two_layer}.  As shown
  Figure~\ref{fig:two_layer}(b) after projection the relative distances within the same cluster
  remain similar while the two clusters become closer to each other.

\paragraph*{\textbf{Clustering with Visual Evidence}}

\begin{figure}[t!]
  \centering
  \includegraphics[width=0.8\linewidth]{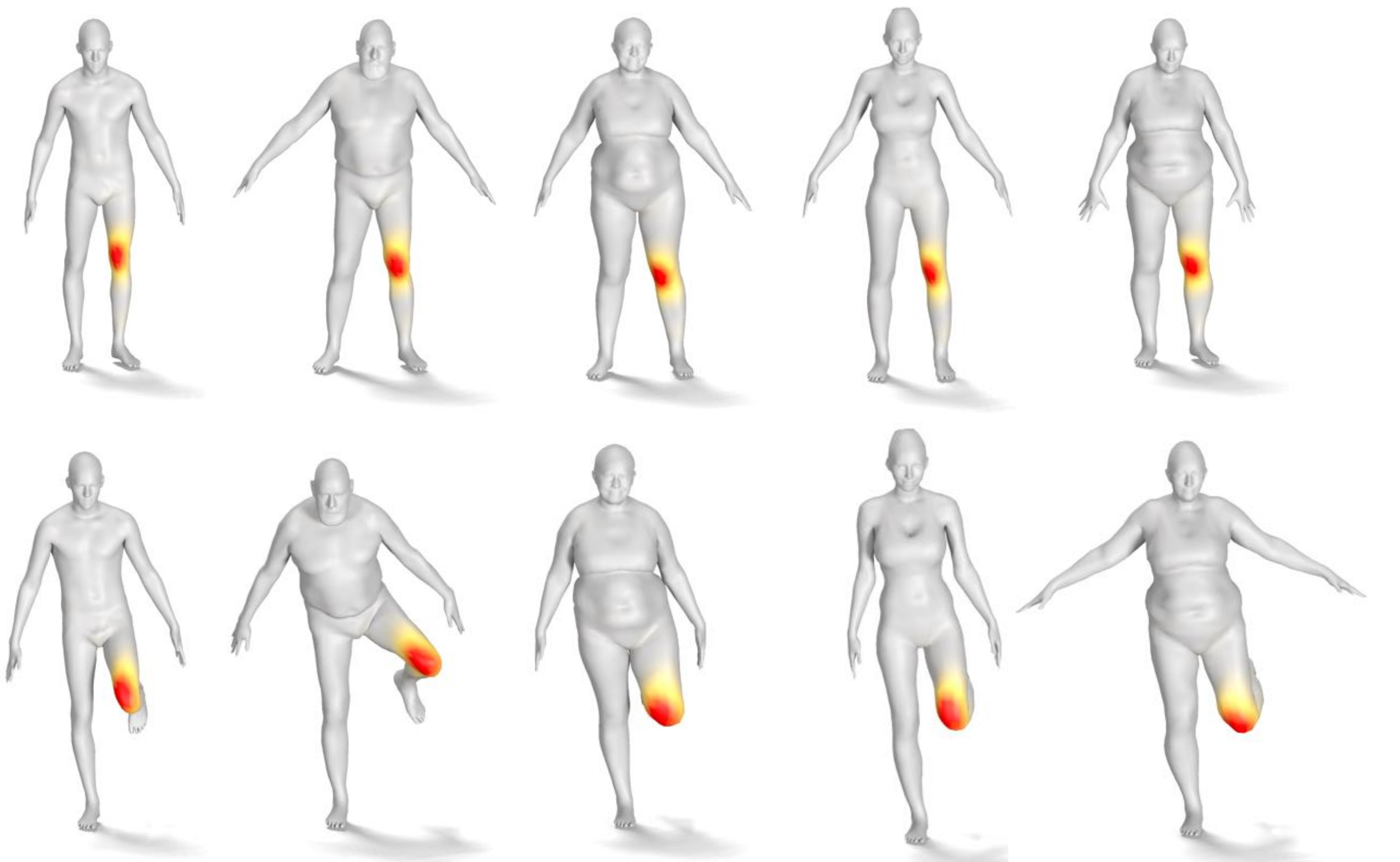}
\caption{\label{fig:compute_knees} The distinctive functions obtained with computed functional maps on the same data set as the one in Figure~\ref{fig:learnsig}. Note that the highlighted regions are comparable.}
\end{figure}

\begin{figure}[ht]
  \centering
  \includegraphics[width=1\linewidth]{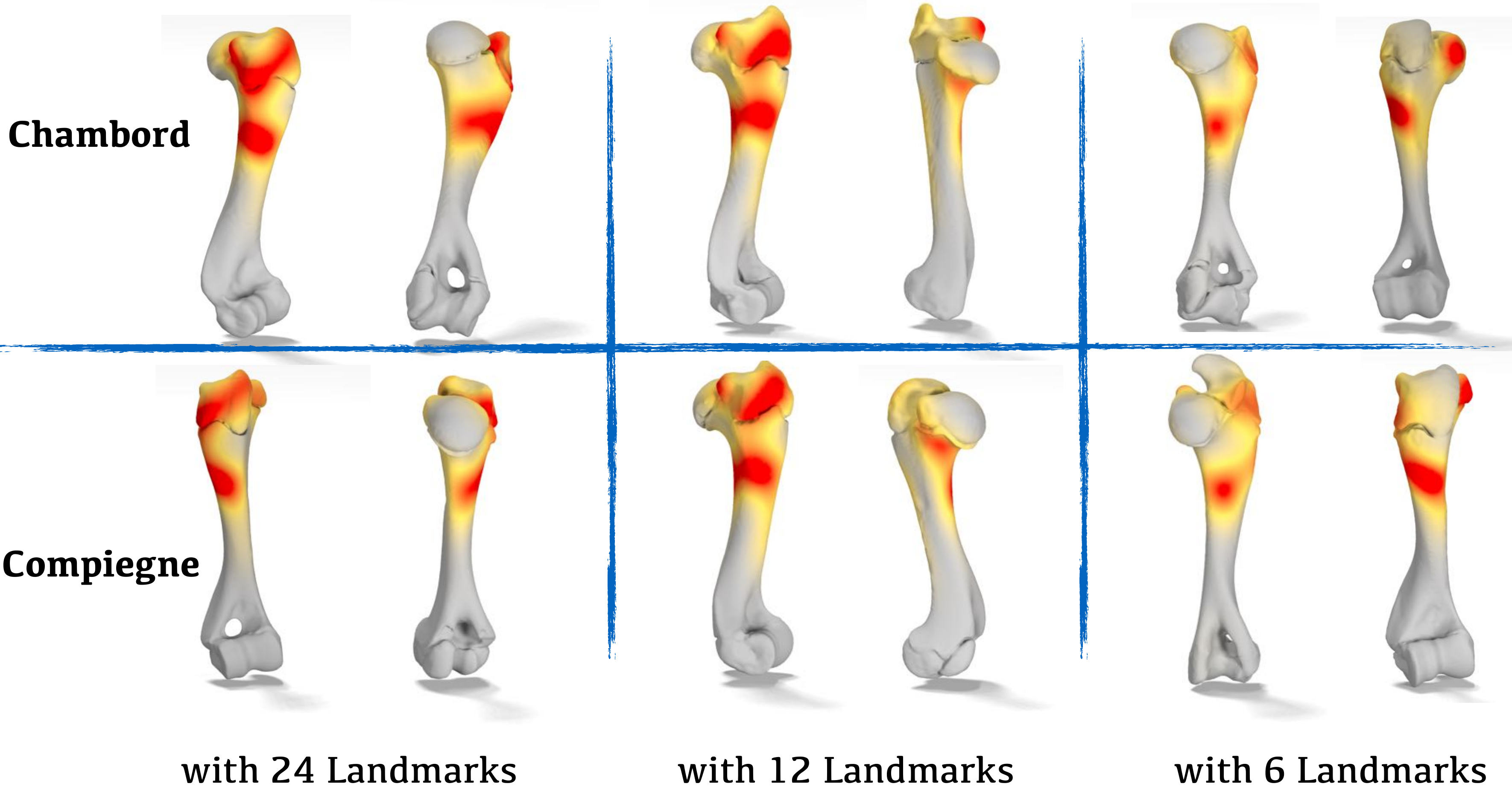}
  \caption{Comparing two sets of bones corresponding to different sub-species of wild boars using computed functional
    maps. The highlighted regions are stable and reveal distinctive yet subtle sub-parts.\label{fig:bones}}
\end{figure}

\begin{figure}[t!]
  \centering
  \includegraphics[width=1\linewidth]{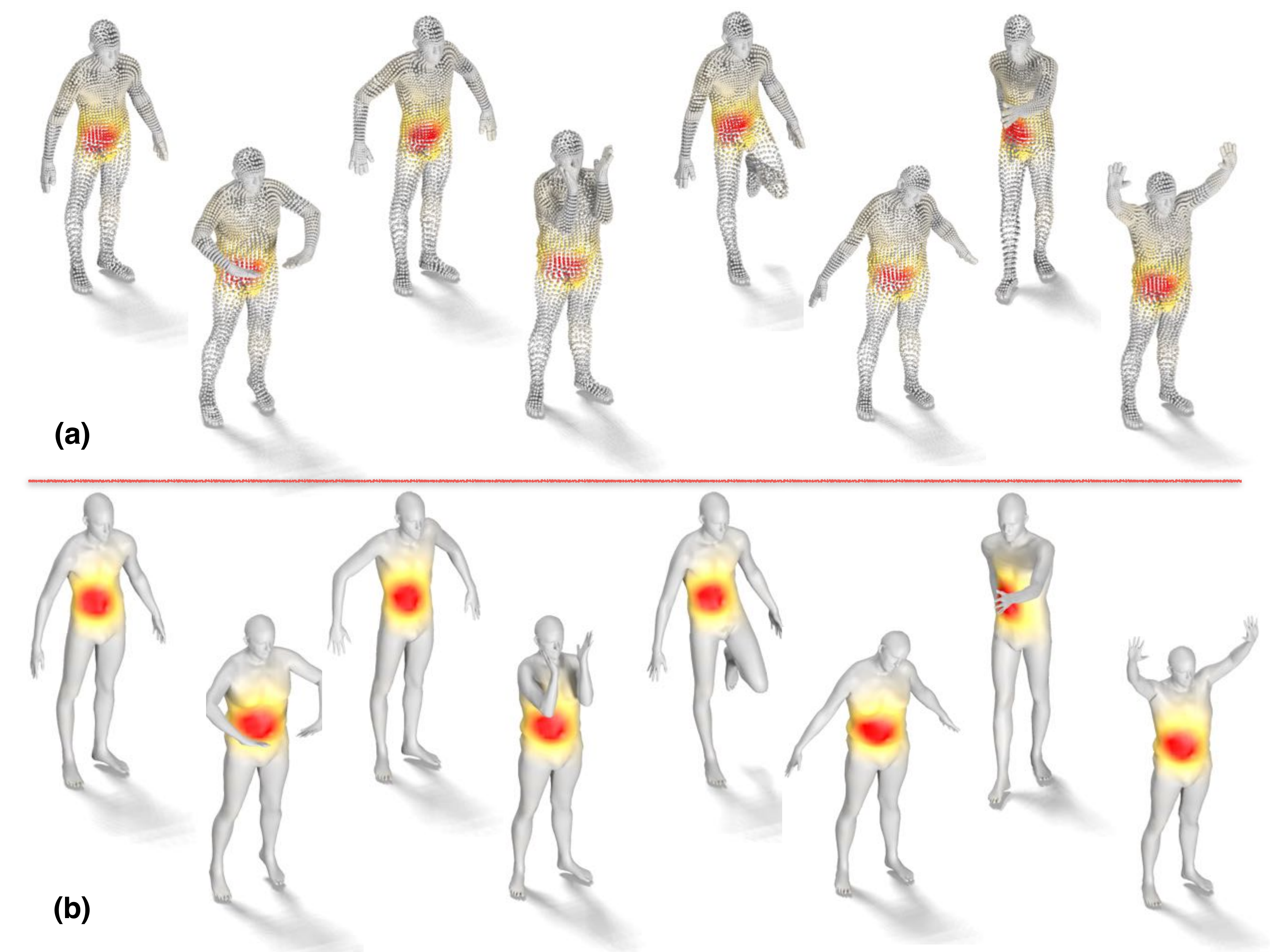}
  \caption{\label{fig:pcd} (a) Comparison between two characters in various poses, represented by point clouds. We plot on them the highlighted function obtained with a \emph{computed} FMN
      among the point clouds. (b) As a baseline, we plot the highlighted function obtained with the
      \emph{ground-truth} FMN on the corresponding meshes.  }
\end{figure}

\begin{figure}[ht]
  \centering
  \includegraphics[width=1\linewidth]{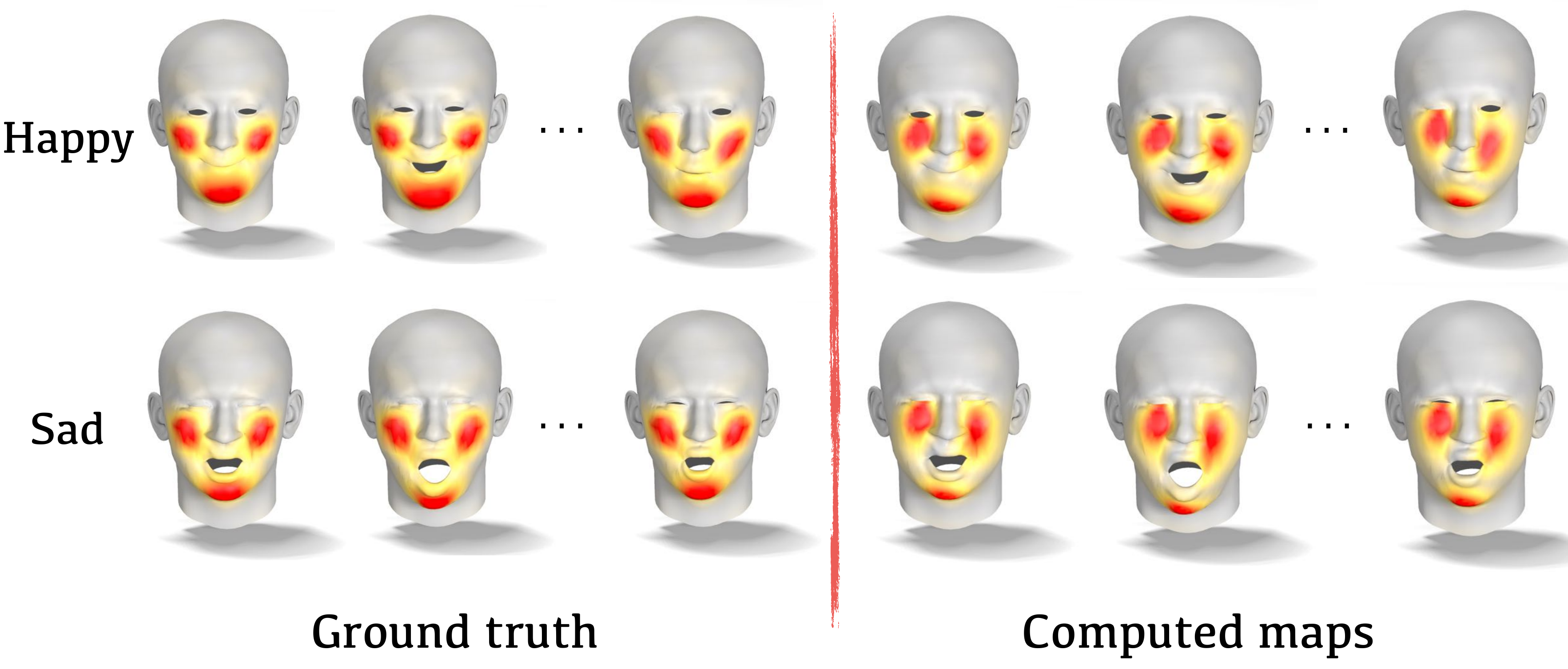}
  \caption{\label{fig:face} Distinctive regions across two sets of expressions (happy on the top, sad on the bottom). Note that the highlighted regions with respect to the computed functional maps are comparable with the ground truth ones.}
\end{figure}

In Figure~\ref{fig:learnsig}, we analyze two clusters of shapes displayed in the two top rows that
represent different characters in two distinct poses. As shown in the first five columns, the
highlighted functions capture the bending knees, which intuitively distinguish the two poses.  In
contrast, both the global variability across the whole collection (the second column from right) and
the ones within each cluster (the right-most column) concentrate on the torso.  We also used \emph{computed functional maps} for detecting the cross-collection variability, and obtained comparable highlighted functions, which are shown on a subset of the shapes in Figure~\ref{fig:compute_knees}. 

We also plot the PCA of the latent shape differences $\{D_i\}_{i = 1}^{20}$ in the bottom of
Figure~\ref{fig:learnsig}, where the blue and red points are mixed, suggesting the dominance of
variability in body type, captured in area-based shape differences.

However, with the highlighted function detected by our approach, expressed through coefficients
$\alpha^*$ in the latent basis, we can separate the shapes by computing $\beta_i = D_i \alpha^*$ and
plotting the PCA of the resulting vectors.  As can be seen in Figure~\ref{fig:learnsig} (bottom,
middle) these vectors separate the two clusters much better.

Furthermore, the same procedure can be performed with partial clustering. Thus, given the cluster
ids of \emph{only} the shapes in the red box, we first computed the optimal distinctive function
$\alpha_{p}^*$ with respect to this subset, and plotted the PCA of $D_i \alpha_{p}^*$ on the latent
shape of the \emph{whole collection}.  The PCA plot of $D_i\alpha_p^*$ suggests that this approach
reveals the correct clusters in a semi-supervised way.

\paragraph*{\textbf{Practical Application in Anatomy}}
The problem of analyzing variability across different classes of 3D objects is well-studied in
computational anatomy, where the classical approach is to first manually establish dense landmark
correspondences and to compute the difference from each object to some pre-computed deformable
template shape. As mentioned above, our approach does not require an actual embedding of a template,
which allows it to handle complex heterogeneous data.

To illustrate this, we compared two sets of bones of two sub-species of wild boars acquired using 3D
scanning techniques. In particular, as input we considered the bone scans with $24$ consistent
handcrafted landmarks and $260$ sliding landmarks~\cite{Gunz2013} on each of shape. We then
estimated the FMN starting with a different number, $(6, 12, 24)$, of the handcrafted landmarks,
using the functional map estimation approach proposed in \cite{Huang2017}. These functional maps
were then used to compute the distinctive regions, as described Section \ref{sec:comparison}. The
corresponding shapes and highlighted functions are shown in Figure~\ref{fig:bones}. Remarkably, our
results are stable with even for a small number of landmarks, and furthermore correspond to
anatomically meaningful shape parts, which in general coincide with the ones detected by the
extrinsic template-based approach, which uses all the $284$ landmarks, and agree with the
functionality explanation for this cross-species variability identified by the domain experts.

\paragraph*{\textbf{Cross-collection Variability across Point Clouds}}
Our framework can also be applied across different modalities. In Figure~\ref{fig:pcd}, we
  compared the shapes on the top row with the ones on the bottom, which clusters correspond to two
  characters in different poses. Therefore the \emph{cross-collection variability} should capture
  the difference in body shapes.  We used the discretization from~\cite{Stability} for computing
  eigenbases on point clouds, and, given a sparse set of ten landmarks across the 8 point clouds, we
  used the adjoint-regularization from~\cite{Huang2017} for estimating the functional maps between
  the shapes.  For comparison, we also computed the cross-collection variability detected with the
  \emph{ground-truth} FMN among the same shapes represented as meshes. Clearly, both of these
  highlight regions are on the torso, reflecting the significant area change, associated with the
  change in body type.

\paragraph*{\textbf{Facial Comparison}} 
We end our demonstration with a comparison between two sets of human faces, which correspond to happy and sad expressions (due to the lack of space, we plot only $6$ faces out of $10$). 
As shown in Figure~\ref{fig:face}, both highlighted functions computed with ground-truth functional maps and the computed ones detect the intuitive cross-collection difference -- the chin and the cheeks. 

%!TEX root = main.tex
\section{Conclusions}
\label{sec:concl}

We have presented a novel approach for representing and analyzing 3D shapes in a context of one or multiple collections. Our construction is based on functional maps network connecting the shapes and a novel analysis that demonstrates that previously used latent functional spaces can both be endowed with a natural geometric structure and provide a basis for representing and comparing shapes in an unbiased way. This leads to Latent Space Shape Differences which represent each shape in the collection as a pair of functional operators, stored as small-sized matrices in practice. This representation has many appealing properties, including invariance to rigid motions as well as full intrinsic informativeness that permits reconstruction. We have demonstrated their use in extracting and highlighting variability of interest in a set of shapes, while also suppressing other variability that we regard as nuisance (and which may in fact manifest in larger geometric deformations). We believe that this highly nuanced understanding of shape distortions and variability is important for many applications in engineering, biology, and medicine. Moreover, we showed that the matrix form of our representation makes it suitable for learning algorithms and the use of CNNs in particular, for both regression and reconstruction.

We also note that the matrix nature of our representation makes it a different mathematical object from the usual latent codes used in machine learning that are invariably points in high-dimensional Euclidean spaces. While point-based representations usually lead to quite limited set of operations (typically, interpolations and vector-based analogies), our difference matrices reflect the internal structure of the shapes and enable, for example, localized shape analogies.

\bibliographystyle{ACM-Reference-Format}
\bibliography{shapemaps-bibliography}

%%% -*-BibTeX-*-
%%% Do NOT edit. File created by BibTeX with style
%%% ACM-Reference-Format-Journals [18-Jan-2012].

\begin{thebibliography}{55}

%%% ====================================================================
%%% NOTE TO THE USER: you can override these defaults by providing
%%% customized versions of any of these macros before the \bibliography
%%% command.  Each of them MUST provide its own final punctuation,
%%% except for \shownote{}, \showDOI{}, and \showURL{}.  The latter two
%%% do not use final punctuation, in order to avoid confusing it with
%%% the Web address.
%%%
%%% To suppress output of a particular field, define its macro to expand
%%% to an empty string, or better, \unskip, like this:
%%%
%%% \newcommand{\showDOI}[1]{\unskip}   % LaTeX syntax
%%%
%%% \def \showDOI #1{\unskip}           % plain TeX syntax
%%%
%%% ====================================================================

\ifx \showCODEN    \undefined \def \showCODEN     #1{\unskip}     \fi
\ifx \showDOI      \undefined \def \showDOI       #1{#1}\fi
\ifx \showISBNx    \undefined \def \showISBNx     #1{\unskip}     \fi
\ifx \showISBNxiii \undefined \def \showISBNxiii  #1{\unskip}     \fi
\ifx \showISSN     \undefined \def \showISSN      #1{\unskip}     \fi
\ifx \showLCCN     \undefined \def \showLCCN      #1{\unskip}     \fi
\ifx \shownote     \undefined \def \shownote      #1{#1}          \fi
\ifx \showarticletitle \undefined \def \showarticletitle #1{#1}   \fi
\ifx \showURL      \undefined \def \showURL       {\relax}        \fi
% The following commands are used for tagged output and should be
% invisible to TeX
\providecommand\bibfield[2]{#2}
\providecommand\bibinfo[2]{#2}
\providecommand\natexlab[1]{#1}
\providecommand\showeprint[2][]{arXiv:#2}

\bibitem[\protect\citeauthoryear{Achlioptas, Diamanti, Mitliagkas, and
  Guibas}{Achlioptas et~al\mbox{.}}{2018}]%
        {achlioptas2017latent_pc}
\bibfield{author}{\bibinfo{person}{Panos Achlioptas}, \bibinfo{person}{Olga
  Diamanti}, \bibinfo{person}{Ioannis Mitliagkas}, {and}
  \bibinfo{person}{Leonidas~J Guibas}.} \bibinfo{year}{2018}\natexlab{}.
\newblock \showarticletitle{Learning Representations and Generative Models For
  3D Point Clouds}.
\newblock \bibinfo{journal}{{\em Proceedings of the 35th International
  Conference on Machine Learning\/}} (\bibinfo{year}{2018}).
\newblock


\bibitem[\protect\citeauthoryear{Allen, Curless, and Popovi{\'c}}{Allen
  et~al\mbox{.}}{2003}]%
        {allen2003space}
\bibfield{author}{\bibinfo{person}{Brett Allen}, \bibinfo{person}{Brian
  Curless}, {and} \bibinfo{person}{Zoran Popovi{\'c}}.}
  \bibinfo{year}{2003}\natexlab{}.
\newblock \showarticletitle{The space of human body shapes: reconstruction and
  parameterization from range scans}. In \bibinfo{booktitle}{{\em ACM
  transactions on graphics (TOG)}}, Vol.~\bibinfo{volume}{22}. ACM,
  \bibinfo{pages}{587--594}.
\newblock


\bibitem[\protect\citeauthoryear{Anguelov, Srinivasan, Koller, Thrun, Rodgers,
  and Davis}{Anguelov et~al\mbox{.}}{2005}]%
        {scape}
\bibfield{author}{\bibinfo{person}{Dragomir Anguelov}, \bibinfo{person}{Praveen
  Srinivasan}, \bibinfo{person}{Daphne Koller}, \bibinfo{person}{Sebastian
  Thrun}, \bibinfo{person}{Jim Rodgers}, {and} \bibinfo{person}{James Davis}.}
  \bibinfo{year}{2005}\natexlab{}.
\newblock \showarticletitle{{SCAPE}: {S}hape {C}ompletion and {A}nimation of
  {P}eople}. In \bibinfo{booktitle}{{\em ACM Transactions on Graphics (TOG)}},
  Vol.~\bibinfo{volume}{24}. ACM, \bibinfo{pages}{408--416}.
\newblock


\bibitem[\protect\citeauthoryear{Blanz and Vetter}{Blanz and Vetter}{1999}]%
        {blanz1999morphable}
\bibfield{author}{\bibinfo{person}{Volker Blanz} {and} \bibinfo{person}{Thomas
  Vetter}.} \bibinfo{year}{1999}\natexlab{}.
\newblock \showarticletitle{A morphable model for the synthesis of 3D faces}.
  In \bibinfo{booktitle}{{\em Proceedings of the 26th annual conference on
  Computer graphics and interactive techniques}}. ACM Press/Addison-Wesley
  Publishing Co., \bibinfo{pages}{187--194}.
\newblock


\bibitem[\protect\citeauthoryear{Bogo, Romero, Loper, and Black}{Bogo
  et~al\mbox{.}}{2014}]%
        {faust}
\bibfield{author}{\bibinfo{person}{Federica Bogo}, \bibinfo{person}{Javier
  Romero}, \bibinfo{person}{Matthew Loper}, {and} \bibinfo{person}{Michael~J
  Black}.} \bibinfo{year}{2014}\natexlab{}.
\newblock \showarticletitle{{FAUST}: {D}ataset and {E}valuation for 3D {M}esh
  {R}egistration}. In \bibinfo{booktitle}{{\em Proc. CVPR}}.
  \bibinfo{pages}{3794--3801}.
\newblock


\bibitem[\protect\citeauthoryear{Bogo, Romero, Pons-Moll, and Black}{Bogo
  et~al\mbox{.}}{2017}]%
        {dfaust}
\bibfield{author}{\bibinfo{person}{Federica Bogo}, \bibinfo{person}{Javier
  Romero}, \bibinfo{person}{Gerard Pons-Moll}, {and}
  \bibinfo{person}{Michael~J. Black}.} \bibinfo{year}{2017}\natexlab{}.
\newblock \showarticletitle{Dynamic {FAUST}: {R}egistering Human Bodies in
  Motion}. In \bibinfo{booktitle}{{\em IEEE Conf. on Computer Vision and
  Pattern Recognition (CVPR)}}.
\newblock


\bibitem[\protect\citeauthoryear{Boscaini, Eynard, Kourounis, and
  Bronstein}{Boscaini et~al\mbox{.}}{2015}]%
        {boscaini2015shape}
\bibfield{author}{\bibinfo{person}{Davide Boscaini}, \bibinfo{person}{Davide
  Eynard}, \bibinfo{person}{Drosos Kourounis}, {and} \bibinfo{person}{Michael~M
  Bronstein}.} \bibinfo{year}{2015}\natexlab{}.
\newblock \showarticletitle{Shape-from-Operator: Recovering Shapes from
  Intrinsic Operators}. In \bibinfo{booktitle}{{\em Computer Graphics Forum}},
  Vol.~\bibinfo{volume}{34}. Wiley Online Library, \bibinfo{pages}{265--274}.
\newblock


\bibitem[\protect\citeauthoryear{Boscaini, Masci, Rodol{\`a}, and
  Bronstein}{Boscaini et~al\mbox{.}}{2016}]%
        {boscaini2016learning}
\bibfield{author}{\bibinfo{person}{Davide Boscaini}, \bibinfo{person}{Jonathan
  Masci}, \bibinfo{person}{Emanuele Rodol{\`a}}, {and} \bibinfo{person}{Michael
  Bronstein}.} \bibinfo{year}{2016}\natexlab{}.
\newblock \showarticletitle{Learning shape correspondence with anisotropic
  convolutional neural networks}. In \bibinfo{booktitle}{{\em Advances in
  Neural Information Processing Systems}}. \bibinfo{pages}{3189--3197}.
\newblock


\bibitem[\protect\citeauthoryear{Bronstein, Bruna, LeCun, Szlam, and
  Vandergheynst}{Bronstein et~al\mbox{.}}{2017}]%
        {bronstein2017geometric}
\bibfield{author}{\bibinfo{person}{Michael~M Bronstein}, \bibinfo{person}{Joan
  Bruna}, \bibinfo{person}{Yann LeCun}, \bibinfo{person}{Arthur Szlam}, {and}
  \bibinfo{person}{Pierre Vandergheynst}.} \bibinfo{year}{2017}\natexlab{}.
\newblock \showarticletitle{Geometric deep learning: going beyond euclidean
  data}.
\newblock \bibinfo{journal}{{\em IEEE Signal Processing Magazine\/}}
  \bibinfo{volume}{34}, \bibinfo{number}{4} (\bibinfo{year}{2017}),
  \bibinfo{pages}{18--42}.
\newblock


\bibitem[\protect\citeauthoryear{Chen, Wang, Li, Su, Wang, Tu, Lischinski,
  Cohen-Or, and Chen}{Chen et~al\mbox{.}}{2015}]%
        {Deep3DPose}
\bibfield{author}{\bibinfo{person}{Wenzheng Chen}, \bibinfo{person}{Huan Wang},
  \bibinfo{person}{Yangyan Li}, \bibinfo{person}{Hao Su},
  \bibinfo{person}{Zhenhua Wang}, \bibinfo{person}{Changhe Tu},
  \bibinfo{person}{Dani Lischinski}, \bibinfo{person}{Daniel Cohen-Or}, {and}
  \bibinfo{person}{Baoquan Chen}.} \bibinfo{year}{2015}\natexlab{}.
\newblock \showarticletitle{Synthesizing Training Images for Boosting Human 3D
  Pose Estimation}. In \bibinfo{booktitle}{{\em 3D Vision (3DV)}}.
\newblock
\showDOI{%
\url{https://doi.org/chen1474147/Deep3DPose}}


\bibitem[\protect\citeauthoryear{Corman, Solomon, Ben-Chen, Guibas, and
  Ovsjanikov}{Corman et~al\mbox{.}}{2017}]%
        {Corman2017}
\bibfield{author}{\bibinfo{person}{Etienne Corman}, \bibinfo{person}{Justin
  Solomon}, \bibinfo{person}{Mirela Ben-Chen}, \bibinfo{person}{Leonidas
  Guibas}, {and} \bibinfo{person}{Maks Ovsjanikov}.}
  \bibinfo{year}{2017}\natexlab{}.
\newblock \showarticletitle{Functional Characterization of Intrinsic and
  Extrinsic Geometry}.
\newblock \bibinfo{journal}{{\em ACM Trans. Graph.\/}} \bibinfo{volume}{36},
  \bibinfo{number}{2}, Article \bibinfo{articleno}{14} (\bibinfo{date}{March}
  \bibinfo{year}{2017}), \bibinfo{numpages}{17}~pages.
\newblock
\showISSN{0730-0301}


\bibitem[\protect\citeauthoryear{Fan, Su, and Guibas}{Fan
  et~al\mbox{.}}{2016}]%
        {fan2016_a}
\bibfield{author}{\bibinfo{person}{Haoqiang Fan}, \bibinfo{person}{Hao Su},
  {and} \bibinfo{person}{Leonidas~J. Guibas}.} \bibinfo{year}{2016}\natexlab{}.
\newblock \showarticletitle{A Point Set Generation Network for 3D Object
  Reconstruction from a Single Image}.
\newblock \bibinfo{journal}{{\em CoRR\/}}  \bibinfo{volume}{abs/1612.00603}
  (\bibinfo{year}{2016}).
\newblock


\bibitem[\protect\citeauthoryear{Girdhar, Fouhey, Rodriguez, and Gupta}{Girdhar
  et~al\mbox{.}}{2016}]%
        {girdhar1eccv}
\bibfield{author}{\bibinfo{person}{Rohit Girdhar}, \bibinfo{person}{David~F.
  Fouhey}, \bibinfo{person}{Mikel Rodriguez}, {and} \bibinfo{person}{Abhinav
  Gupta}.} \bibinfo{year}{2016}\natexlab{}.
\newblock \bibinfo{booktitle}{{\em Learning a Predictable and Generative Vector
  Representation for Objects}}.
\newblock \bibinfo{publisher}{Springer International Publishing},
  \bibinfo{address}{Cham}, \bibinfo{pages}{484--499}.
\newblock
\showISBNx{978-3-319-46466-4}


\bibitem[\protect\citeauthoryear{Grenander and Miller}{Grenander and
  Miller}{1998}]%
        {grenander1998computational}
\bibfield{author}{\bibinfo{person}{Ulf Grenander} {and}
  \bibinfo{person}{Michael~I Miller}.} \bibinfo{year}{1998}\natexlab{}.
\newblock \showarticletitle{Computational anatomy: An emerging discipline}.
\newblock \bibinfo{journal}{{\em Quarterly of applied mathematics\/}}
  \bibinfo{volume}{56}, \bibinfo{number}{4} (\bibinfo{year}{1998}),
  \bibinfo{pages}{617--694}.
\newblock


\bibitem[\protect\citeauthoryear{Gunz and Mitteroecker}{Gunz and
  Mitteroecker}{2013}]%
        {Gunz2013}
\bibfield{author}{\bibinfo{person}{Philipp Gunz} {and} \bibinfo{person}{Philipp
  Mitteroecker}.} \bibinfo{year}{2013}\natexlab{}.
\newblock \showarticletitle{Semilandmarks: a method for quantifying curves and
  surfaces}.
\newblock \bibinfo{journal}{{\em Hystrix, the Italian Journal of Mammalogy\/}}
  \bibinfo{volume}{24}, \bibinfo{number}{1} (\bibinfo{year}{2013}),
  \bibinfo{pages}{103--109}.
\newblock
\showISSN{0394-1914}
\showDOI{%
\url{https://doi.org/10.4404/hystrix-24.1-6292}}


\bibitem[\protect\citeauthoryear{Hasler, Stoll, Sunkel, Rosenhahn, and
  Seidel}{Hasler et~al\mbox{.}}{2009}]%
        {hasler09}
\bibfield{author}{\bibinfo{person}{Nils Hasler}, \bibinfo{person}{Carsten
  Stoll}, \bibinfo{person}{Martin Sunkel}, \bibinfo{person}{Bodo Rosenhahn},
  {and} \bibinfo{person}{H-P Seidel}.} \bibinfo{year}{2009}\natexlab{}.
\newblock \showarticletitle{A {S}tatistical {M}odel of {H}uman {P}ose and
  {B}ody {S}hape}. In \bibinfo{booktitle}{{\em Computer Graphics Forum}},
  Vol.~\bibinfo{volume}{28}. \bibinfo{pages}{337--346}.
\newblock


\bibitem[\protect\citeauthoryear{Huang, Wang, and Guibas}{Huang
  et~al\mbox{.}}{2014}]%
        {huang2014functional}
\bibfield{author}{\bibinfo{person}{Qixing Huang}, \bibinfo{person}{Fan Wang},
  {and} \bibinfo{person}{Leonidas Guibas}.} \bibinfo{year}{2014}\natexlab{}.
\newblock \showarticletitle{Functional map networks for analyzing and exploring
  large shape collections}.
\newblock \bibinfo{journal}{{\em ACM Transactions on Graphics (TOG)\/}}
  \bibinfo{volume}{33}, \bibinfo{number}{4} (\bibinfo{year}{2014}),
  \bibinfo{pages}{36}.
\newblock


\bibitem[\protect\citeauthoryear{Huang, Chazal, and Ovsjanikov}{Huang
  et~al\mbox{.}}{2017}]%
        {Stability}
\bibfield{author}{\bibinfo{person}{Ruqi Huang}, \bibinfo{person}{Frederic
  Chazal}, {and} \bibinfo{person}{Maks Ovsjanikov}.}
  \bibinfo{year}{2017}\natexlab{}.
\newblock \showarticletitle{On the Stability of Functional Maps and Shape
  Difference Operators}.
\newblock  \bibinfo{volume}{37}, \bibinfo{number}{1} (\bibinfo{year}{2017}).
\newblock


\bibitem[\protect\citeauthoryear{Huang and Ovsjanikov}{Huang and
  Ovsjanikov}{2017}]%
        {Huang2017}
\bibfield{author}{\bibinfo{person}{Ruqi Huang} {and} \bibinfo{person}{Maks
  Ovsjanikov}.} \bibinfo{year}{2017}\natexlab{}.
\newblock \showarticletitle{Adjoint Map Representation for Shape Analysis and
  Matching}. In \bibinfo{booktitle}{{\em Proc. SGP}},
  Vol.~\bibinfo{volume}{36}.
\newblock


\bibitem[\protect\citeauthoryear{Joshi, Davis, Jomier, and Gerig}{Joshi
  et~al\mbox{.}}{2004}]%
        {joshi2004unbiased}
\bibfield{author}{\bibinfo{person}{Sarang Joshi}, \bibinfo{person}{Brad Davis},
  \bibinfo{person}{Matthieu Jomier}, {and} \bibinfo{person}{Guido Gerig}.}
  \bibinfo{year}{2004}\natexlab{}.
\newblock \showarticletitle{Unbiased diffeomorphic atlas construction for
  computational anatomy}.
\newblock \bibinfo{journal}{{\em NeuroImage\/}}  \bibinfo{volume}{23}
  (\bibinfo{year}{2004}), \bibinfo{pages}{S151--S160}.
\newblock


\bibitem[\protect\citeauthoryear{Kendall}{Kendall}{1989}]%
        {kendall1989survey}
\bibfield{author}{\bibinfo{person}{David~G Kendall}.}
  \bibinfo{year}{1989}\natexlab{}.
\newblock \showarticletitle{A survey of the statistical theory of shape}.
\newblock \bibinfo{journal}{{\it Statist. Sci.}} (\bibinfo{year}{1989}),
  \bibinfo{pages}{87--99}.
\newblock


\bibitem[\protect\citeauthoryear{Kim, Li, Mitra, Chaudhuri, DiVerdi, and
  Funkhouser}{Kim et~al\mbox{.}}{2013}]%
        {kim2013learning}
\bibfield{author}{\bibinfo{person}{Vladimir~G Kim}, \bibinfo{person}{Wilmot
  Li}, \bibinfo{person}{Niloy~J Mitra}, \bibinfo{person}{Siddhartha Chaudhuri},
  \bibinfo{person}{Stephen DiVerdi}, {and} \bibinfo{person}{Thomas
  Funkhouser}.} \bibinfo{year}{2013}\natexlab{}.
\newblock \showarticletitle{Learning part-based templates from large
  collections of 3D shapes}.
\newblock \bibinfo{journal}{{\em ACM Transactions on Graphics (TOG)\/}}
  \bibinfo{volume}{32}, \bibinfo{number}{4} (\bibinfo{year}{2013}),
  \bibinfo{pages}{70}.
\newblock


\bibitem[\protect\citeauthoryear{Kim, Li, Mitra, DiVerdi, and Funkhouser}{Kim
  et~al\mbox{.}}{2012}]%
        {kim2012exploring}
\bibfield{author}{\bibinfo{person}{Vladimir~G Kim}, \bibinfo{person}{Wilmot
  Li}, \bibinfo{person}{Niloy~J Mitra}, \bibinfo{person}{Stephen DiVerdi},
  {and} \bibinfo{person}{Thomas Funkhouser}.} \bibinfo{year}{2012}\natexlab{}.
\newblock \showarticletitle{Exploring collections of 3D models using fuzzy
  correspondences}.
\newblock \bibinfo{journal}{{\em ACM Transactions on Graphics (TOG)\/}}
  \bibinfo{volume}{31}, \bibinfo{number}{4} (\bibinfo{year}{2012}),
  \bibinfo{pages}{54}.
\newblock


\bibitem[\protect\citeauthoryear{Kingma and Ba}{Kingma and Ba}{2014}]%
        {Adam}
\bibfield{author}{\bibinfo{person}{Diederik~P. Kingma} {and}
  \bibinfo{person}{Jimmy Ba}.} \bibinfo{year}{2014}\natexlab{}.
\newblock \showarticletitle{Adam: {A} Method for Stochastic Optimization}.
\newblock \bibinfo{journal}{{\em CoRR\/}}  \bibinfo{volume}{abs/1412.6980}
  (\bibinfo{year}{2014}).
\newblock
\showeprint[arxiv]{1412.6980}


\bibitem[\protect\citeauthoryear{Kleiman, van Kaick, Sorkine-Hornung, and
  Cohen-Or}{Kleiman et~al\mbox{.}}{2015}]%
        {kleiman2015shed}
\bibfield{author}{\bibinfo{person}{Yanir Kleiman}, \bibinfo{person}{Oliver van
  Kaick}, \bibinfo{person}{Olga Sorkine-Hornung}, {and} \bibinfo{person}{Daniel
  Cohen-Or}.} \bibinfo{year}{2015}\natexlab{}.
\newblock \showarticletitle{SHED: shape edit distance for fine-grained shape
  similarity}.
\newblock \bibinfo{journal}{{\em ACM Transactions on Graphics (TOG)\/}}
  \bibinfo{volume}{34}, \bibinfo{number}{6} (\bibinfo{year}{2015}),
  \bibinfo{pages}{235}.
\newblock


\bibitem[\protect\citeauthoryear{Kovnatsky, Bronstein, Bronstein, Glashoff, and
  Kimmel}{Kovnatsky et~al\mbox{.}}{2013}]%
        {kovnatsky2013coupled}
\bibfield{author}{\bibinfo{person}{Artiom Kovnatsky},
  \bibinfo{person}{Michael~M Bronstein}, \bibinfo{person}{Alexander~M
  Bronstein}, \bibinfo{person}{Klaus Glashoff}, {and} \bibinfo{person}{Ron
  Kimmel}.} \bibinfo{year}{2013}\natexlab{}.
\newblock \showarticletitle{Coupled quasi-harmonic bases}. In
  \bibinfo{booktitle}{{\em Computer Graphics Forum}},
  Vol.~\bibinfo{volume}{32}. Wiley Online Library, \bibinfo{pages}{439--448}.
\newblock


\bibitem[\protect\citeauthoryear{L\"{a}hner, Vestner, Boyarski, Litany,
  Slossberg, Remez, Rodol`a, Bronstein, Bronstein, Kimmel, and
  Cremers}{L\"{a}hner et~al\mbox{.}}{2017}]%
        {kernel17}
\bibfield{author}{\bibinfo{person}{Z. L\"{a}hner}, \bibinfo{person}{M.
  Vestner}, \bibinfo{person}{A. Boyarski}, \bibinfo{person}{O. Litany},
  \bibinfo{person}{R. Slossberg}, \bibinfo{person}{T. Remez},
  \bibinfo{person}{E. Rodol`a}, \bibinfo{person}{A.~M. Bronstein},
  \bibinfo{person}{M.~M. Bronstein}, \bibinfo{person}{R. Kimmel}, {and}
  \bibinfo{person}{D. Cremers}.} \bibinfo{year}{2017}\natexlab{}.
\newblock \showarticletitle{Efficient Deformable Shape Correspondence via
  Kernel Matching}.
\newblock \bibinfo{journal}{{\em arXiv preprint 1707.08991\/}}
  (\bibinfo{year}{2017}).
\newblock


\bibitem[\protect\citeauthoryear{Li, Xu, Chaudhuri, Yumer, Zhang, and
  Guibas}{Li et~al\mbox{.}}{2017}]%
        {LiXCYZG17}
\bibfield{author}{\bibinfo{person}{Jun Li}, \bibinfo{person}{Kai Xu},
  \bibinfo{person}{Siddhartha Chaudhuri}, \bibinfo{person}{Ersin Yumer},
  \bibinfo{person}{Hao Zhang}, {and} \bibinfo{person}{Leonidas~J. Guibas}.}
  \bibinfo{year}{2017}\natexlab{}.
\newblock \showarticletitle{{GRASS:} Generative Recursive Autoencoders for
  Shape Structures}.
\newblock \bibinfo{journal}{{\em CoRR\/}}  \bibinfo{volume}{abs/1705.02090}
  (\bibinfo{year}{2017}).
\newblock


\bibitem[\protect\citeauthoryear{Maron, Galun, Aigerman, Trope, Dym, Yumer,
  KIM, and Lipman}{Maron et~al\mbox{.}}{2017}]%
        {maron2017convolutional}
\bibfield{author}{\bibinfo{person}{Haggai Maron}, \bibinfo{person}{Meirav
  Galun}, \bibinfo{person}{Noam Aigerman}, \bibinfo{person}{Miri Trope},
  \bibinfo{person}{Nadav Dym}, \bibinfo{person}{Ersin Yumer},
  \bibinfo{person}{VLADIMIR~G KIM}, {and} \bibinfo{person}{Yaron Lipman}.}
  \bibinfo{year}{2017}\natexlab{}.
\newblock \showarticletitle{Convolutional Neural Networks on Surfaces via
  Seamless Toric Covers}.
\newblock


\bibitem[\protect\citeauthoryear{Masci, Boscaini, Bronstein, and
  Vandergheynst}{Masci et~al\mbox{.}}{2015}]%
        {masci2015geodesic}
\bibfield{author}{\bibinfo{person}{Jonathan Masci}, \bibinfo{person}{Davide
  Boscaini}, \bibinfo{person}{Michael Bronstein}, {and} \bibinfo{person}{Pierre
  Vandergheynst}.} \bibinfo{year}{2015}\natexlab{}.
\newblock \showarticletitle{Geodesic convolutional neural networks on
  riemannian manifolds}. In \bibinfo{booktitle}{{\em Proc. ICCV workshops}}.
  \bibinfo{pages}{37--45}.
\newblock


\bibitem[\protect\citeauthoryear{Maturana and Scherer}{Maturana and
  Scherer}{2015}]%
        {maturana2015voxnet}
\bibfield{author}{\bibinfo{person}{Daniel Maturana} {and}
  \bibinfo{person}{Sebastian Scherer}.} \bibinfo{year}{2015}\natexlab{}.
\newblock \showarticletitle{Voxnet: A 3d convolutional neural network for
  real-time object recognition}. In \bibinfo{booktitle}{{\em Intelligent Robots
  and Systems (IROS), 2015 IEEE/RSJ International Conference on}}. IEEE,
  \bibinfo{pages}{922--928}.
\newblock


\bibitem[\protect\citeauthoryear{Meyer, Desbrun, Schr{\"o}der, and Barr}{Meyer
  et~al\mbox{.}}{2003}]%
        {meyer2003discrete}
\bibfield{author}{\bibinfo{person}{Mark Meyer}, \bibinfo{person}{Mathieu
  Desbrun}, \bibinfo{person}{Peter Schr{\"o}der}, {and} \bibinfo{person}{Alan~H
  Barr}.} \bibinfo{year}{2003}\natexlab{}.
\newblock \showarticletitle{Discrete differential-geometry operators for
  triangulated 2-manifolds}.
\newblock In \bibinfo{booktitle}{{\em Visualization and mathematics III}}.
  \bibinfo{publisher}{Springer}, \bibinfo{pages}{35--57}.
\newblock


\bibitem[\protect\citeauthoryear{Mikolov, Sutskever, Chen, Corrado, and
  Dean}{Mikolov et~al\mbox{.}}{2013}]%
        {DBLP:journals/corr/MikolovSCCD13}
\bibfield{author}{\bibinfo{person}{Tomas Mikolov}, \bibinfo{person}{Ilya
  Sutskever}, \bibinfo{person}{Kai Chen}, \bibinfo{person}{Greg Corrado}, {and}
  \bibinfo{person}{Jeffrey Dean}.} \bibinfo{year}{2013}\natexlab{}.
\newblock \showarticletitle{Distributed Representations of Words and Phrases
  and their Compositionality}.
\newblock \bibinfo{journal}{{\em CoRR\/}}  \bibinfo{volume}{abs/1310.4546}
  (\bibinfo{year}{2013}).
\newblock
\showeprint[arxiv]{1310.4546}


\bibitem[\protect\citeauthoryear{Ovsjanikov, Ben-Chen, Chazal, and
  Guibas}{Ovsjanikov et~al\mbox{.}}{2013}]%
        {Ovsjanikov2013}
\bibfield{author}{\bibinfo{person}{M. Ovsjanikov}, \bibinfo{person}{M.
  Ben-Chen}, \bibinfo{person}{F. Chazal}, {and} \bibinfo{person}{L. Guibas}.}
  \bibinfo{year}{2013}\natexlab{}.
\newblock \showarticletitle{{Analysis and visualization of maps between
  shapes}}.
\newblock \bibinfo{journal}{{\em Computer Graphics Forum\/}}
  \bibinfo{volume}{32}, \bibinfo{number}{6} (\bibinfo{year}{2013}),
  \bibinfo{pages}{135--145}.
\newblock


\bibitem[\protect\citeauthoryear{Ovsjanikov, Ben-Chen, Solomon, Butscher, and
  Guibas}{Ovsjanikov et~al\mbox{.}}{2012}]%
        {Functional}
\bibfield{author}{\bibinfo{person}{Maks Ovsjanikov}, \bibinfo{person}{Mirela
  Ben-Chen}, \bibinfo{person}{Justin Solomon}, \bibinfo{person}{Adrian
  Butscher}, {and} \bibinfo{person}{Leonidas Guibas}.}
  \bibinfo{year}{2012}\natexlab{}.
\newblock \showarticletitle{{F}unctional {M}aps: {A} {F}lexible
  {R}epresentation of {M}aps {B}etween {S}hapes}.
\newblock \bibinfo{journal}{{\em ACM Transactions on Graphics (TOG)\/}}
  \bibinfo{volume}{31}, \bibinfo{number}{4} (\bibinfo{year}{2012}),
  \bibinfo{pages}{30}.
\newblock


\bibitem[\protect\citeauthoryear{Ovsjanikov, Corman, Bronstein, Rodol{\`a},
  Ben-Chen, Guibas, Chazal, and Bronstein}{Ovsjanikov et~al\mbox{.}}{2017}]%
        {ovsjanikov2017computing}
\bibfield{author}{\bibinfo{person}{Maks Ovsjanikov}, \bibinfo{person}{Etienne
  Corman}, \bibinfo{person}{Michael Bronstein}, \bibinfo{person}{Emanuele
  Rodol{\`a}}, \bibinfo{person}{Mirela Ben-Chen}, \bibinfo{person}{Leonidas
  Guibas}, \bibinfo{person}{Frederic Chazal}, {and} \bibinfo{person}{Alex
  Bronstein}.} \bibinfo{year}{2017}\natexlab{}.
\newblock \showarticletitle{Computing and processing correspondences with
  functional maps}. In \bibinfo{booktitle}{{\em ACM SIGGRAPH 2017 Courses}}.
  ACM, \bibinfo{pages}{5}.
\newblock


\bibitem[\protect\citeauthoryear{Ovsjanikov, Li, Guibas, and Mitra}{Ovsjanikov
  et~al\mbox{.}}{2011}]%
        {ovsjanikov2011exploration}
\bibfield{author}{\bibinfo{person}{Maks Ovsjanikov}, \bibinfo{person}{Wilmot
  Li}, \bibinfo{person}{Leonidas Guibas}, {and} \bibinfo{person}{Niloy~J
  Mitra}.} \bibinfo{year}{2011}\natexlab{}.
\newblock \showarticletitle{Exploration of continuous variability in
  collections of 3D shapes}.
\newblock \bibinfo{journal}{{\em ACM Transactions on Graphics (TOG)\/}}
  \bibinfo{volume}{30}, \bibinfo{number}{4} (\bibinfo{year}{2011}),
  \bibinfo{pages}{33}.
\newblock


\bibitem[\protect\citeauthoryear{Qi, Su, Mo, and Guibas}{Qi
  et~al\mbox{.}}{2016}]%
        {qi2016_pointnet}
\bibfield{author}{\bibinfo{person}{Charles~Ruizhongtai Qi},
  \bibinfo{person}{Hao Su}, \bibinfo{person}{Kaichun Mo}, {and}
  \bibinfo{person}{Leonidas~J. Guibas}.} \bibinfo{year}{2016}\natexlab{}.
\newblock \showarticletitle{PointNet: deep learning on point sets for 3D
  classification and segmentation}.
\newblock \bibinfo{journal}{{\em CoRR\/}}  \bibinfo{volume}{abs/1612.00593}
  (\bibinfo{year}{2016}).
\newblock


\bibitem[\protect\citeauthoryear{Reuter, Wolter, and Peinecke}{Reuter
  et~al\mbox{.}}{2006}]%
        {Reuter2006}
\bibfield{author}{\bibinfo{person}{Martin Reuter}, \bibinfo{person}{Franz-Erich
  Wolter}, {and} \bibinfo{person}{Niklas Peinecke}.}
  \bibinfo{year}{2006}\natexlab{}.
\newblock \showarticletitle{Laplace-Beltrami Spectra As 'Shape-DNA' of Surfaces
  and Solids}.
\newblock \bibinfo{journal}{{\em Comput. Aided Des.\/}} \bibinfo{volume}{38},
  \bibinfo{number}{4} (\bibinfo{date}{April} \bibinfo{year}{2006}),
  \bibinfo{pages}{342--366}.
\newblock
\showISSN{0010-4485}


\bibitem[\protect\citeauthoryear{Rustamov, Ovsjanikov, Azencot, Ben-Chen,
  Chazal, and Guibas}{Rustamov et~al\mbox{.}}{2013}]%
        {Rustamov2013}
\bibfield{author}{\bibinfo{person}{Raif~M. Rustamov}, \bibinfo{person}{Maks
  Ovsjanikov}, \bibinfo{person}{Omri Azencot}, \bibinfo{person}{Mirela
  Ben-Chen}, \bibinfo{person}{Fr{\'{e}}d{\'{e}}ric Chazal}, {and}
  \bibinfo{person}{Leonidas Guibas}.} \bibinfo{year}{2013}\natexlab{}.
\newblock \showarticletitle{{Map-based exploration of intrinsic shape
  differences and variability}}.
\newblock \bibinfo{journal}{{\em ACM Transactions on Graphics\/}}
  \bibinfo{volume}{32}, \bibinfo{number}{4} (\bibinfo{year}{2013}),
  \bibinfo{pages}{1}.
\newblock


\bibitem[\protect\citeauthoryear{Sinha, Bai, and Ramani}{Sinha
  et~al\mbox{.}}{2016}]%
        {sinha2016deep}
\bibfield{author}{\bibinfo{person}{Ayan Sinha}, \bibinfo{person}{Jing Bai},
  {and} \bibinfo{person}{Karthik Ramani}.} \bibinfo{year}{2016}\natexlab{}.
\newblock \showarticletitle{Deep learning 3d shape surfaces using geometry
  images}. In \bibinfo{booktitle}{{\em European Conference on Computer
  Vision}}. Springer, \bibinfo{pages}{223--240}.
\newblock


\bibitem[\protect\citeauthoryear{Solomon, Nguyen, Butscher, Ben-Chen, and
  Guibas}{Solomon et~al\mbox{.}}{2012}]%
        {solomon2012soft}
\bibfield{author}{\bibinfo{person}{Justin Solomon}, \bibinfo{person}{Andy
  Nguyen}, \bibinfo{person}{Adrian Butscher}, \bibinfo{person}{Mirela
  Ben-Chen}, {and} \bibinfo{person}{Leonidas Guibas}.}
  \bibinfo{year}{2012}\natexlab{}.
\newblock \showarticletitle{Soft maps between surfaces}. In
  \bibinfo{booktitle}{{\em Computer Graphics Forum}},
  Vol.~\bibinfo{volume}{31}. Wiley Online Library, \bibinfo{pages}{1617--1626}.
\newblock


\bibitem[\protect\citeauthoryear{Su, Maji, Kalogerakis, and Learned-Miller}{Su
  et~al\mbox{.}}{2015}]%
        {Su_volumetric}
\bibfield{author}{\bibinfo{person}{Hang Su}, \bibinfo{person}{Subhransu Maji},
  \bibinfo{person}{Evangelos Kalogerakis}, {and} \bibinfo{person}{Erik
  Learned-Miller}.} \bibinfo{year}{2015}\natexlab{}.
\newblock \showarticletitle{Multi-view Convolutional Neural Networks for 3D
  Shape Recognition}. In \bibinfo{booktitle}{{\em Proceedings of the 2015 IEEE
  International Conference on Computer Vision (ICCV)}} {\em
  (\bibinfo{series}{ICCV '15})}.
\newblock


\bibitem[\protect\citeauthoryear{Thompson et~al\mbox{.}}{Thompson
  et~al\mbox{.}}{1942}]%
        {thompson1942growth}
\bibfield{author}{\bibinfo{person}{Darcy~Wentworth Thompson} {et~al\mbox{.}}}
  \bibinfo{year}{1942}\natexlab{}.
\newblock \showarticletitle{On growth and form.}
\newblock \bibinfo{journal}{{\em On growth and form.\/}}
  (\bibinfo{year}{1942}).
\newblock


\bibitem[\protect\citeauthoryear{Tong, Zhou, Liu, Pan, and Yan}{Tong
  et~al\mbox{.}}{2012}]%
        {tong2012scanning}
\bibfield{author}{\bibinfo{person}{Jing Tong}, \bibinfo{person}{Jin Zhou},
  \bibinfo{person}{Ligang Liu}, \bibinfo{person}{Zhigeng Pan}, {and}
  \bibinfo{person}{Hao Yan}.} \bibinfo{year}{2012}\natexlab{}.
\newblock \showarticletitle{Scanning 3d full human bodies using kinects}.
\newblock \bibinfo{journal}{{\em IEEE transactions on visualization and
  computer graphics\/}} \bibinfo{volume}{18}, \bibinfo{number}{4}
  (\bibinfo{year}{2012}), \bibinfo{pages}{643--650}.
\newblock


\bibitem[\protect\citeauthoryear{Wand, Adams, Ovsjanikov, Berner, Bokeloh,
  Jenke, Guibas, Seidel, and Schilling}{Wand et~al\mbox{.}}{2009}]%
        {wand2009efficient}
\bibfield{author}{\bibinfo{person}{Michael Wand}, \bibinfo{person}{Bart Adams},
  \bibinfo{person}{Maksim Ovsjanikov}, \bibinfo{person}{Alexander Berner},
  \bibinfo{person}{Martin Bokeloh}, \bibinfo{person}{Philipp Jenke},
  \bibinfo{person}{Leonidas Guibas}, \bibinfo{person}{Hans-Peter Seidel}, {and}
  \bibinfo{person}{Andreas Schilling}.} \bibinfo{year}{2009}\natexlab{}.
\newblock \showarticletitle{Efficient reconstruction of nonrigid shape and
  motion from real-time 3D scanner data}.
\newblock \bibinfo{journal}{{\em ACM Transactions on Graphics (TOG)\/}}
  \bibinfo{volume}{28}, \bibinfo{number}{2} (\bibinfo{year}{2009}),
  \bibinfo{pages}{15}.
\newblock


\bibitem[\protect\citeauthoryear{Wand, Jenke, Huang, Bokeloh, Guibas, and
  Schilling}{Wand et~al\mbox{.}}{2007}]%
        {wand2007}
\bibfield{author}{\bibinfo{person}{Michael Wand}, \bibinfo{person}{Philipp
  Jenke}, \bibinfo{person}{Qi-Xing Huang}, \bibinfo{person}{Martin Bokeloh},
  \bibinfo{person}{Leonidas Guibas}, {and} \bibinfo{person}{Andreas
  Schilling}.} \bibinfo{year}{2007}\natexlab{}.
\newblock \showarticletitle{Reconstruction of Deforming Geometry from
  Time-Varying Point Clouds}. In \bibinfo{booktitle}{{\em Proc. SGP}}.
  \bibinfo{pages}{49--58}.
\newblock


\bibitem[\protect\citeauthoryear{Wang, Huang, and Guibas}{Wang
  et~al\mbox{.}}{2013}]%
        {Wang2013}
\bibfield{author}{\bibinfo{person}{Fan Wang}, \bibinfo{person}{Qixing Huang},
  {and} \bibinfo{person}{Leonidas~J. Guibas}.} \bibinfo{year}{2013}\natexlab{}.
\newblock \showarticletitle{{Image co-segmentation via consistent functional
  maps}}. In \bibinfo{booktitle}{{\em Proceedings of the IEEE International
  Conference on Computer Vision}}. \bibinfo{pages}{849--856}.
\newblock


\bibitem[\protect\citeauthoryear{Wang, Huang, Ovsjanikov, and Guibas}{Wang
  et~al\mbox{.}}{2014}]%
        {wang2014unsupervised}
\bibfield{author}{\bibinfo{person}{Fan Wang}, \bibinfo{person}{Qixing Huang},
  \bibinfo{person}{Maks Ovsjanikov}, {and} \bibinfo{person}{Leonidas~J
  Guibas}.} \bibinfo{year}{2014}\natexlab{}.
\newblock \showarticletitle{Unsupervised multi-class joint image segmentation}.
  In \bibinfo{booktitle}{{\em Proceedings of the IEEE Conference on Computer
  Vision and Pattern Recognition}}. \bibinfo{pages}{3142--3149}.
\newblock


\bibitem[\protect\citeauthoryear{Wang and Singer}{Wang and Singer}{2013}]%
        {wang2013exact}
\bibfield{author}{\bibinfo{person}{Lanhui Wang} {and} \bibinfo{person}{Amit
  Singer}.} \bibinfo{year}{2013}\natexlab{}.
\newblock \showarticletitle{Exact and stable recovery of rotations for robust
  synchronization}.
\newblock \bibinfo{journal}{{\em Information and Inference: A Journal of the
  IMA\/}} \bibinfo{volume}{2}, \bibinfo{number}{2} (\bibinfo{year}{2013}),
  \bibinfo{pages}{145--193}.
\newblock


\bibitem[\protect\citeauthoryear{Wang, Liu, Guo, Sun, and Tong}{Wang
  et~al\mbox{.}}{2017}]%
        {wang2017cnn}
\bibfield{author}{\bibinfo{person}{Peng-Shuai Wang}, \bibinfo{person}{Yang
  Liu}, \bibinfo{person}{Yu-Xiao Guo}, \bibinfo{person}{Chun-Yu Sun}, {and}
  \bibinfo{person}{Xin Tong}.} \bibinfo{year}{2017}\natexlab{}.
\newblock \showarticletitle{O-cnn: Octree-based convolutional neural networks
  for 3d shape analysis}.
\newblock \bibinfo{journal}{{\em ACM Transactions on Graphics (TOG)\/}}
  \bibinfo{volume}{36}, \bibinfo{number}{4} (\bibinfo{year}{2017}),
  \bibinfo{pages}{72}.
\newblock


\bibitem[\protect\citeauthoryear{Wang, Asafi, van Kaick, Zhang, Cohen-Or, and
  Chen}{Wang et~al\mbox{.}}{2012}]%
        {wang2012active}
\bibfield{author}{\bibinfo{person}{Yunhai Wang}, \bibinfo{person}{Shmulik
  Asafi}, \bibinfo{person}{Oliver van Kaick}, \bibinfo{person}{Hao Zhang},
  \bibinfo{person}{Daniel Cohen-Or}, {and} \bibinfo{person}{Baoquan Chen}.}
  \bibinfo{year}{2012}\natexlab{}.
\newblock \showarticletitle{Active co-analysis of a set of shapes}.
\newblock \bibinfo{journal}{{\em ACM Transactions on Graphics (TOG)\/}}
  \bibinfo{volume}{31}, \bibinfo{number}{6} (\bibinfo{year}{2012}),
  \bibinfo{pages}{165}.
\newblock


\bibitem[\protect\citeauthoryear{Wu, Zhang, Xue, Freeman, and Tenenbaum}{Wu
  et~al\mbox{.}}{2016}]%
        {3dgan}
\bibfield{author}{\bibinfo{person}{Jiajun Wu}, \bibinfo{person}{Chengkai
  Zhang}, \bibinfo{person}{Tianfan Xue}, \bibinfo{person}{Bill Freeman}, {and}
  \bibinfo{person}{Josh Tenenbaum}.} \bibinfo{year}{2016}\natexlab{}.
\newblock \showarticletitle{Learning a Probabilistic Latent Space of Object
  Shapes via 3D Generative-Adversarial Modeling}. In \bibinfo{booktitle}{{\em
  Proc. NIPS}}. \bibinfo{pages}{82--90}.
\newblock


\bibitem[\protect\citeauthoryear{Younes}{Younes}{2010}]%
        {younes2010shapes}
\bibfield{author}{\bibinfo{person}{Laurent Younes}.}
  \bibinfo{year}{2010}\natexlab{}.
\newblock \bibinfo{booktitle}{{\em Shapes and diffeomorphisms}}.
  Vol.~\bibinfo{volume}{171}.
\newblock \bibinfo{publisher}{Springer Science \& Business Media}.
\newblock


\bibitem[\protect\citeauthoryear{Zeng, Guo, Luo, and Gu}{Zeng
  et~al\mbox{.}}{2012}]%
        {zeng2012discrete}
\bibfield{author}{\bibinfo{person}{Wei Zeng}, \bibinfo{person}{Ren Guo},
  \bibinfo{person}{Feng Luo}, {and} \bibinfo{person}{Xianfeng Gu}.}
  \bibinfo{year}{2012}\natexlab{}.
\newblock \showarticletitle{Discrete heat kernel determines discrete Riemannian
  metric}.
\newblock \bibinfo{journal}{{\em Graphical Models\/}} \bibinfo{volume}{74},
  \bibinfo{number}{4} (\bibinfo{year}{2012}), \bibinfo{pages}{121--129}.
\newblock


\end{thebibliography}

%!TEX root = main.tex

\appendix
\section{Technical Details}

\paragraph*{Proof of Theorem~\ref{thm:latentshape}}
\begin{proof}
First note that $\tP$ is well-defined since by consistency, $\Phi_j Y_j = \Phi_j C_{ij} Y_i = \Phi_i Y_i$. 
The regularization constraint $\sum_i Y_i^T Y_i = Id$ therefore implies $\sum_i (\Phi_i M_i \tP)^T \Phi_i M_i \tP = \tP^T (\sum_i M_i) \tP = Id$. 

Now let $E = \sum_i Y_i^T \Lambda_i Y_i$ to be a diagonal matrix (implicitly corresponds to the eigenvalues of the latent shape). 
Note that $\Lambda_i$ is a non-negative diagonal matrix, thus $E$ admits an eigen-decomposition $E U = U \Lambda_0$ and we let $\Phi_0 = \tP U$.
Direct computation yields that $\Phi_0^T (\sum_i M_i) \Phi_0 = Id$, and $\Phi_0^T (\sum_i L_i) \Phi_0 = \Lambda_0$. 
Thus it follows from $\Phi_0^T M \Phi_0 \Lambda_0 = \Lambda_0 = \Phi_0^T L \Phi_0$ that $L \Phi_0 = M \Phi_0 \Lambda_0$. On the other hand, it is easy to verify that the eigenfunctions of $(L, M)$ satisfies the the consistency constraint and the normalization, therefore they are equivalent.
\end{proof}

\paragraph*{Proof of Lemma~\ref{lem:delta_energy}}
\begin{proof}
We first prove that:
\[\Vert D_i - D_j\Vert_{\mbox{Fro}}^2 - \Vert P_i(F) - P_j(F)\Vert_{\mbox{Fro}}^2 = \Vert (D_i - D_j) - (P_i(F) - P_j(F))\Vert_{\mbox{Fro}}^2. \]
It is easy to verify that $(F F^T)^2 = (F F^T)$, since $F^T F = Id$. 
In other words, $F F^T$ is a projection operator, then so is $Id - F F^T$. 
For the sake of simplicity, we denote in the following $D_i - D_j$ and $Id - F F^T$ by $\Delta, K$ respectively. 
Obviously $\Delta, K$ are both symmetric matrices, and $K^2 = K$. 
Then the above equivalence can be re-rewritten as
\[\mbox{Trace}(\Delta^T \Delta) - \mbox{Trace}(K^T \Delta^T \Delta K) = \mbox{Trace} [(\Delta - \Delta K)^T(\Delta - \Delta K)], \]
which amounts to $\mbox{Trace}(K^T \Delta^T \Delta K) = \mbox{Trace}(K^T \Delta^T \Delta)$.

Finally, the equivalence follows from 
\begin{align*}
	\mbox{Trace}(K^T \Delta^T \Delta K) &= \mbox{Trace}(\Delta^T \Delta K K^T) \\
   & = \mbox{Trace}(\Delta^T \Delta K^2) = \mbox{Trace}(\Delta^T \Delta K).
\end{align*}

Finally, the difference is equal to 
\[
	\Vert \Delta K\Vert_{\mbox{Fro}}^2  =  \mbox{Trace}(K\Delta^2 K)= \mbox{Trace}(F^T \Delta^2 F(F^T F)) = \mbox{Trace}(F^T \Delta^2 F)
\]
\end{proof}

\paragraph*{Connection between our Method and the framework of~\cite{Huang2017}}
Our formulation constructs a linear combination of terms $H^{D}(i,j) = (D_k - D_l)^2$,$\sum_{i, j} w^D_{ij}H^D(i, j)$, where $w^D_{ij} = w^D_{ji}$, and then computing the eigenvectors associated with the largest eigenvalues of it 
In essence, the distortion energy constructed in~\cite{Huang2017} is similarly composed of a set of terms in the form $H^X(i,j) = (X_{ij}Y_i - Y_j)^T(X_{ij}Y_i - Y_j)$, where $X_{ij}$ is the adjoint functional map from $S_i$ to $S_j$, and $Y_i$ is the latent basis on $S_i$. 

Our main observation is that, in the case of area-based operators, under the same condition as of Theorem~\ref{thm:latentshape}, $H^X(k, l) H^D(i, j) = H^D(i, j) H^X(k, l), \forall i, j, k,l$. 
A consequence of the above argument is that when both the spectra of $\sum_{i, j}w^D_{ij} H^D(i, j)$ and $\sum_{i, j}w^X_{ij} H^X(i, j)$ have no repeating eigenvalues, then their eigenvectors are identical. 

We provide a sketch proof of this claim. Following the proof of Theorem~\ref{thm:latentshape}, we have $Y_i = \Phi_i^{-1}\Phi_0$, where $\Phi_i$ is the \emph{full} eigenbasis on $S_i$, and $\Phi_0$ is the eigenbasis of the average/latent shape. 
Then we have $D_i = Y_i^T Y_i = \Phi_0^T \Phi_i^{-T} \Phi_i^{-1} \Phi_0 = \Phi_0^T M_i \Phi_0$, which implies $H^D(i, j) = \Phi_0^T (M_i - M_j) M_0 (M_i - M_j) \Phi_0$, where $M_0$ is the measure of the average shape.
Regarding the adjoint case, we similarly have $H^X(k, l) = \Phi_0^T (M_k^2 M_l + M_l -2 M_k)\Phi_0$. Therefore it is easy to verify the commutativity between $H^D(i,j)$ and $H^X(k, l)$.

\section{Neural Network Details}
\subsection{Regression}
\subsubsection{Point-Cloud Architectures}
\label{sec:PC-archs}
We used three configurations for making point-base architectures. In a spirit similar to \cite{qi2016_pointnet} we implemented all (3-layer deep) encoders as $1$-D convolutions with filter size $1$, i.e., treating each point independently. The output of the last encoding layer was further processed by a feature-wise max-pool which was further processed by an FC-ReLU decoder. Table~\ref{table:pc-arch-params} shows the exact number of parameters (columns) in each consecutive layer for the three configurations (rows).

\begin{table}[ht]
    \begin{center}
    \begin{tabular}{c c c}
        \hline
        Version   & Encoder (\# filters) & Decoder (\# Neurons)\\
        \hline
        \hline
        A   &\{32, 64, 64\}    &\{64, 12\} \\
        B   &\{64, 128, 128\}  &\{64, 12\} \\
        C   &\{64, 128, 128\}  &\{64, 128, 12\} \\
        \hline
    \end{tabular}
    \caption{Size of layers in point-based architectures for the versions that formed the  baseline of the regression experiments. The further right a parameter is displayed the  deeper the underlying layer of architecture is.}
    \label{table:pc-arch-params}
    \end{center}
\end{table}

We trained each of these architectures with learning rates of \{0.001, 0.002, 0.005, 0.007, 0.01\}. The learning rate of $0.005$ gave the best performance in the regression experiments.. 

\subsubsection{MLPs}
We used FC-ReLU MLPS for which the last 3 layers had \{50, 100, 12\} neurons respectively. 
The number of neurons of the first layer was calibrated according to the size of the input difference matrix. Table~\ref{table:mlp-arch-params} shows their correspondence.
\begin{table}[ht]    
    \centering
    \begin{tabular}{ c | c c c c c c c}
        \hline
        \# Latent-Bases  & 5   & 10 & 20 & 30 & 40 & 50 \\
        \# Neurons       & 369 & 185& 62 & 29 & 17 & 11 \\
        \hline                     
    \end{tabular}
    \caption{Number of neurons in first layer of MLP-architectures based on the size corresponding to the $\#$Latent-Bases.}
    \label{table:mlp-arch-params}    
\end{table}

\subsubsection{CNNs}
The encoding part of our CNNs was comprised by two convolutional layers leading to a single FC-ReLU layer with $12$ neurons. See Table~\ref{table:cnn-arch-parms-in20} and Table~\ref{table:cnn-arch-parms-in40} for the parameters of the convolutional layers when the input was $20\times20$ difference matrices and $40\times40$, respectively.
\begin{table}[ht]
    \centering
    \begin{tabular}{|c|c|c|c|}
        \hline
        Layer & \# Filters  & Kernel-size & Stride \\
        \hline
        First  & 10 & (2, 2) & 1\\
        \hline
        Second & 10 & (4, 4) & 2\\
        \hline
    \end{tabular}
    \vspace{10pt}
    \caption{CNN parameters with $20\times20$ input.}
    \label{table:cnn-arch-parms-in20}
\end{table}

\begin{table}[ht]
    \centering
    \begin{tabular}{|c|c|c|c|}
        \hline
        Layer & \# Filters  & Kernel-size & Stride \\
        \hline
        First  & 10 & (3, 3) & 2\\
        \hline
        Second & 10 & (4, 4) & 2\\
        \hline
    \end{tabular}
    \vspace{10pt}
    \caption{CNN parameters with $40\times40$ input.}
    \label{table:cnn-arch-parms-in40}
\end{table}

\subsection{Reconstruction Architecture}
\label{sec:recon-arch}
The architecture we used here is inspired by the CNN used for regression. Again, the convolutional part comes with two encoding layers (see Table~\ref{table:cnn-arch-for-recon} for parameters). The decoder is an MLP implemented with FC-ReLU layers of size $\{128, 128, 4096 \times 3\}$.

\begin{table}[ht]
    \centering
    \begin{tabular}{|c|c|c|c|}
        \hline
        Layer & \# Filters  & Kernel-size & Stride \\
        \hline
        First  & 20 & (3, 3) & 2\\
        \hline
        Second & 20 & (6, 6) & 2\\
        \hline
    \end{tabular}
    \vspace{10pt}
    \caption{CNN encoding parameters with $40\times40$ input for the purposed of reconstruction a point-cloud from a difference-matrix.}
    \label{table:cnn-arch-for-recon}
\end{table}

\subsection{Training details}
For training we used stochastic gradient descent with Adam \cite{Adam} ($\beta_1 = 0.9$) and batch-size of $50$ throughout all experiments. Moreover we normalized the differences matrices by subtracting their average wrt. the training split. For the regression task, the networks operating with difference-matrices were trained with a learning rate of $0.007$. In the reconstruction experiments we trained the CNN-architecture for 850 epochs with a learning rate of $0.005$.

\begin{figure}[ht]
    \centering
    \subfloat{\includegraphics[width=70mm, scale=0.7]{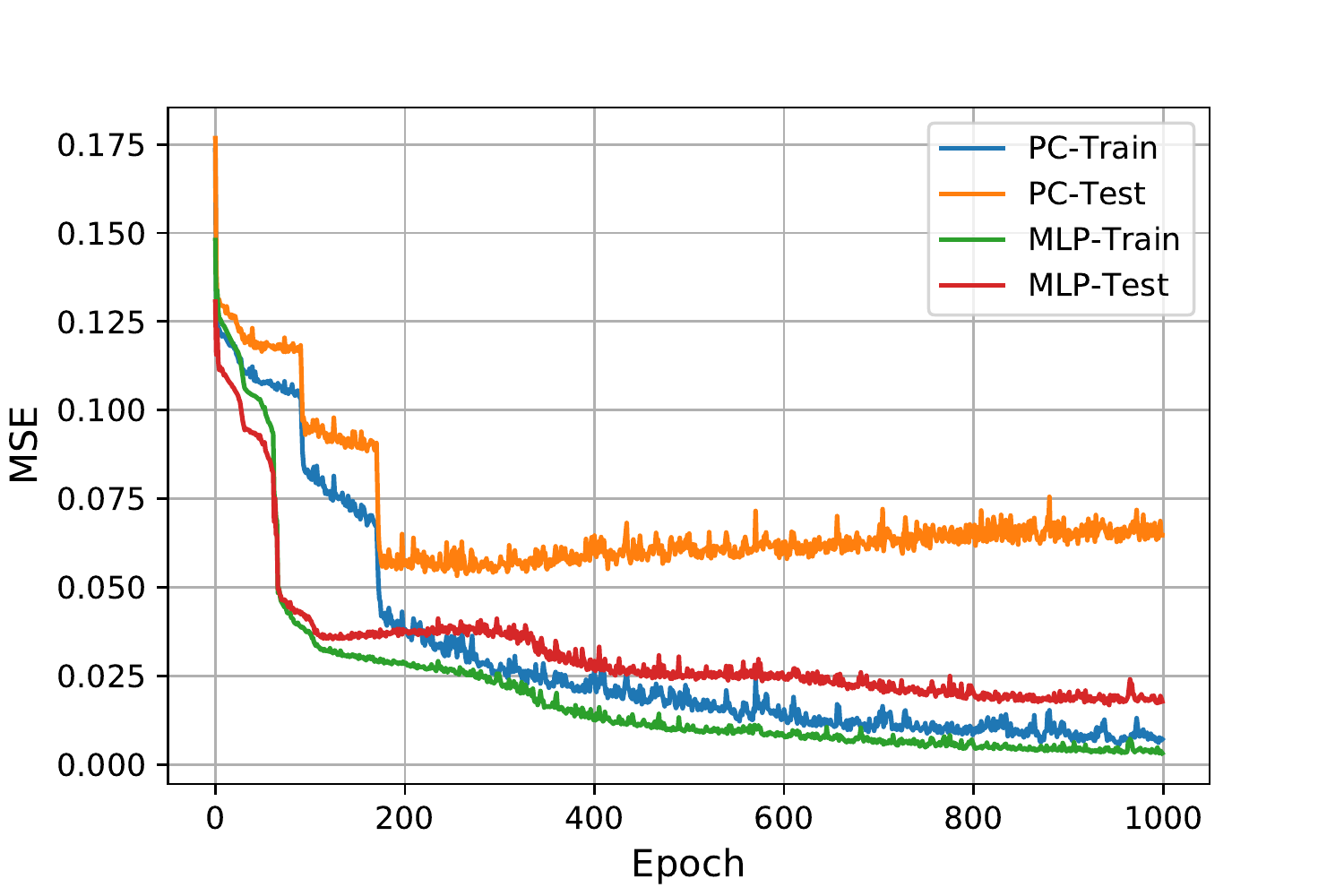} }\
	\subfloat{\includegraphics[width=70mm, scale=0.7]{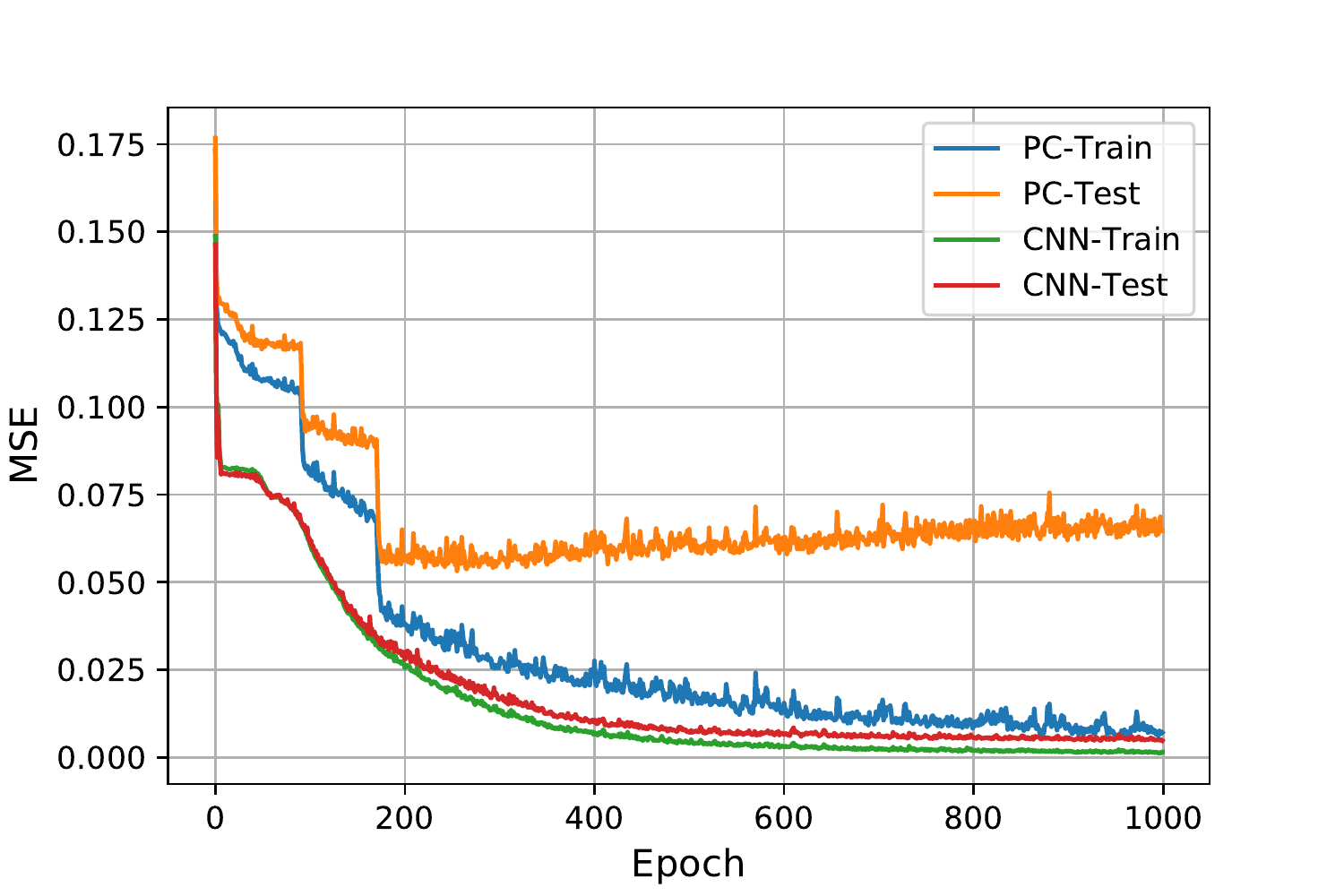} }    
    \caption{Training trends: PC-based Net vs. MLP (top) or CNN(bottom) architectures on $40\times40$ difference maps. The PC-based architecture suffers from a lot of over-fitting while the CNN one enjoys very good generalization error. (Plots based on training with single seed.)} \label{fig:mse-training-trends}
\end{figure}

\end{document}